\documentclass[final,3p,8pt]{elsarticle}

\usepackage{amssymb}
\usepackage{array}
\usepackage{mathrsfs}
\usepackage{amscd}
\usepackage{amsmath}
\usepackage{amsfonts}
\usepackage{amsthm}
\usepackage{psfrag}
\usepackage{textcomp}
\usepackage{bm}
\usepackage{dsfont}
\usepackage{algorithm}
\usepackage{algpseudocode}

\usepackage{amsthm}

 \usepackage{lineno}

\journal{\texttt{arXiv.org}}

\usepackage{amssymb}

\biboptions{square,sort&compress}

\usepackage[figuresright]{rotating}
\usepackage{moreverb}
\usepackage{amssymb}
\usepackage{array}
\usepackage{mathrsfs}
\usepackage{amscd}
\usepackage{amsmath}
\usepackage{amsfonts}
\usepackage{amsthm}
\usepackage{psfrag}
\usepackage{textcomp}
\usepackage{bm}
\usepackage{color}
\usepackage{float}
\usepackage[latin1]{inputenc} 	
\usepackage{afterpage}
\usepackage{tabularx}
\usepackage{booktabs}
\usepackage{romannum}
\usepackage{multirow}
\usepackage[normalem]{ulem} 
\usepackage{cancel}

\begin{document}
	

\begin{frontmatter}




\title{Wave propagation in beams with multiple resonators: conditions for weak scattering and the Born approximation}


\author[upv2,icl]{Mario L\'azaro\corref{cor}}
\ead{malana@upv.es}
\cortext[cor]{Corresponding author. Tel +34 963877000 (Ext. 76732). Fax +34 963877189}


\author[icl]{Richard Wiltshaw}
\author[icl]{Richard V. Craster}
\author[upv2]{Vicent Romero-Garc\'ia}

\address[upv2]{Instituto Universitario de Matem\'atica  Pura y Aplicada, 
	Universitat Polit\`ecnica de Val\`encia, 46022 (Spain)}
\address[icl]{Department of Mathematics, Imperial College London, London, SW7 2AZ, UK}

\begin{abstract}

This  {work reports} the conditions under which weak scattering assumptions can be applied in a beam {loaded by} multiple resonators  {supporting} both longitudinal and flexural waves. The {work} derives the equations of motion of a one-dimensional elastic waveguide with several point resonators by utilizing the Green's matrix approach.  The derivations include any resonator morphology, either with a discrete or continuous distribution of resonances. The method employed is based on applying multiple scattering theory. The response can be expressed as an infinite series whose convergence is closely linked to the scattering intensity provided by the resonators. The convergence conditions are reduced to studying the spectral radius of the scattering matrix. Furthermore, the  {the leading order of the multiple scattering expansion} is associated with the Born approximation. The work also provides approximate expressions for the spectral radius, offering a physical interpretation to the concept of weak scattering in beams. Several numerical examples are presented to validate the proposed methodology.


\end{abstract}


\end{frontmatter}


\section{Introduction}



Wave scattering by an assembling, spatial distribution  of interacting scatterers is a subject of interest across various fields of science and technology. This includes areas like  condensed matter physics, optics, acoustics or mechanics. In general,  incident waves are simultaneously scattered and dissipated by the obstacles, leading to a multiple scattering process that depends on the shape,  material properties, and dimensions of the individual scatterers and on the spatial correlation of the assembling.  

The control of wave scattering {is} a subject of {continuous} discussion in {the community of wave} physics {including} electromagnetism \cite{Pfeiffer-2013}, photonics\cite{ZhangY-2019}, and acoustics \cite{Jimenez-2017a}. However, in recent years, there has been {an increasing interest} on engineered materials designed to manipulate waves {on demand}. Examples such as photonic \cite{Yablonovitch-1987,Sajeev-1987,Joannopoulos-2011} or phononic crystals \cite{Sigalas-1992,MartinezSala-1995,Deymier-2013b}, hyperuniform and stealth materials \cite{Torquato-2002,Torquato-2004,Torquato-2008,Torquato-2016b,Torquato-2015}, along with metamaterials \cite{Engheta-2006,Zhengyou-2000,FangN-2006,Wong-2017}, illustrate {a wide plethora} of many-body systems used to manage the scattering of waves. {The scattered wavefield can be semi-analytically or numerically calculated by several methods as multiple scattering theory \cite{Martin-2006} or finite element methods \cite{Ihlenburg-1998}. However, in some situations the computational cost can be huge and several approximations may be used}. One of them is theBorn approximation {allowing the estimation of} the wavefield under weak scattering conditions. This approach has a long history in quantum mechanics and has recently been used in other fields such as acoustics for the design of new media with stealth or equiluminus properties \cite{RomeroGarcia-2019,RomeroGarcia-2021}.

Research into the scattering of elastic waves has placed significant emphasis on understanding the interactions among point defects within clusters. This includes the influential work of Foldy \cite{Foldy-1945}  and Lax \cite{Lax-1951}, which laid the groundwork for representing defects as sources, whether isotropic or otherwise, with unspecified intensities. Martin \cite{Martin-2006} provides a comprehensive review of scattering theories and their connection to integral equations. Mace provided explicit analytical solutions for the reflection and transmission coefficients of beams, considering both Euler-Bernoulli \cite{Mace-1984} and Timoshenko \cite{Mace-2005} beams, while also accounting for the influence of evanescent modes. Utilizing techniques rooted in the spectral element method has proven to be highly effective in simulating defects and cracks in structures at high frequencies, without the need for computationally intensive finite element models \cite{WangCH-2003,Krawczuk-2003a,Palacz-2005a,TanC-1998,Nieves-2017,Ryue-2011}. The development of phononic metamaterials has garnered significant attention in recent decades. This interest is driven by the potential to manipulate wave behavior through modifications to the medium, achieved via single-frequency oscillators \cite{WangG-2005} or elastic metamaterials with multiple resonators  \cite{Miranda-2019,WangZ-2013,Xiao-2012c,Xiao-2013}. Additionally, novel configurations incorporating attached Rayleigh beams offer the ability to control the response by coupling longitudinal and flexural waves  \cite{Movchan-2022a,Movchan-2022b,Movchan-2023}. In the realm of flexural waves in plates, the introduction of point oscillators necessitates the application of the multiple scattering method in conjunction with the plane wave expansion approach \cite{Torrent-2012,Wiltshaw-2020,Movchan-2017a,Movchan-2018,Ruzzene-2017}. \\

The examination of how variations in medium heterogeneity and property perturbations affect resonances is of significant practical interest in fields like experimental modal analysis and structural health monitoring. Researchers have suggested combining analytical approximations with finite element method-based techniques to analyze vibrating structures with varying properties along their length  \cite{LeeJ-2016,Adamek-2015,Aya-2012,ZhangK-2017}. Moreover, for assessing transmission and reflection resulting from the introduction of multiple oscillators, methods integrating analytical approximations with finite element analysis have been proposed \cite{WuJ-1998,Brennan-1999,Mace-2007,TanC-1998}.  Additionally, models incorporating heterogeneity generated by internal cracks using rotational springs have been suggested \cite{Krawczuk-1994a,Krawczuk-1996,Krawczuk-2001a,Krawczuk-2003b}. In the mechanics of nanostructures, models based on property contrast have been put forth to analyze the response of structures with multiple cracks distributed along their length and their impact on modal parameters. This applies to both flexural  \cite{Loghmani-2018a} and longitudinal waves \cite{Loghmani-2018b}.\\

This paper outlines a precise methodology for deriving the one-dimensional flexural and longitudinal wavefield within a system of resonators of any kind. It introduces a formulation based on a Green's matrix explicitly derived to encompass the influence of all scatterer resonances within a single matrix. By applying multiple scattering theory to beams, the problem can be explicitly solved, and conditions for approximating the solution iteratively are derived. The series obtained include the Born approximation as the zero-iteration case. These developments reveal that the weak scattering condition is closely tied to the spectral radius of the scattering matrix, providing a means to assess the quality of the Born approximation in the context of flexural-longitudinal waves in beams. The paper also presents numerical examples to validate this approximation and establish its connection with the theoretical derivations.

\section{Elastic waveguides with multiple attached resonators}

\begin{figure}[h]%
	\begin{center}
		\begin{tabular}{c}
			\includegraphics[width=17cm]{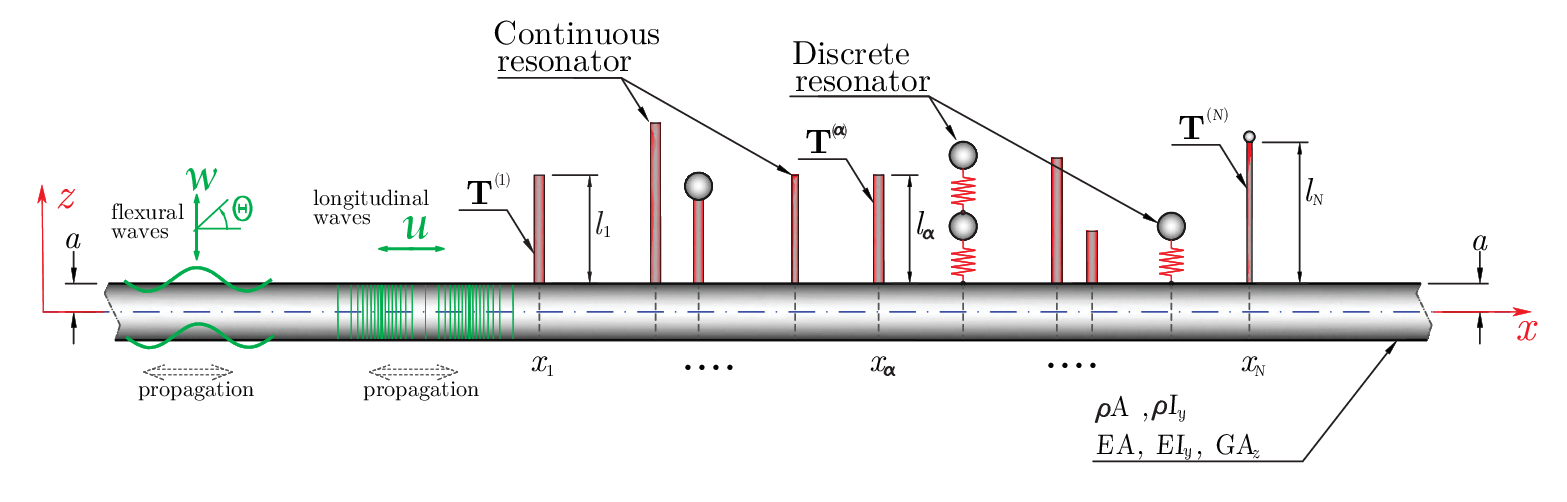} 
		\end{tabular}			
		\caption{Beam-rod general configuration with $N$ scatterers-resonators of any type}%
		\label{fig01}%
	\end{center}
\end{figure}

We consider longitudinal and flexural waves propagating through a straight infinite beam-rod waveguide. We assume also that the reference axis $x$ coincides with the centroid of the cross-section, so no coupling between both type of waves are presented if the medium is homogeneous. In this paper,  an arrangement of $N$ scattterers with internal resonances and distributed along the beam at locations $x_1,\ldots,x_N$ will be also considered.  We do not make any restriction except that the motion of both beam and resonators is restricted to the $xz$--plane (considered as a symmetry plane, see Fig.~\ref{fig01}). Either discrete or continuous resonators can be considered, depending on whether they have a finite or infinite number of resonance frequencies, respectively. Longitudinal and transversal vibrations of the scatterer induce forces and moments back to the main beam via the attachment point.  scattering consequently the incident wave.\\

The homogeneous medium will be modeled using classical approach for longitudinal waves and shear deformations with rotational inertia for flexural waves (Timoshenko beam). Thus, the equations of motion are 

\begin{figure}[h]%
	\begin{center}
		\begin{tabular}{c}
			\includegraphics[width=13cm]{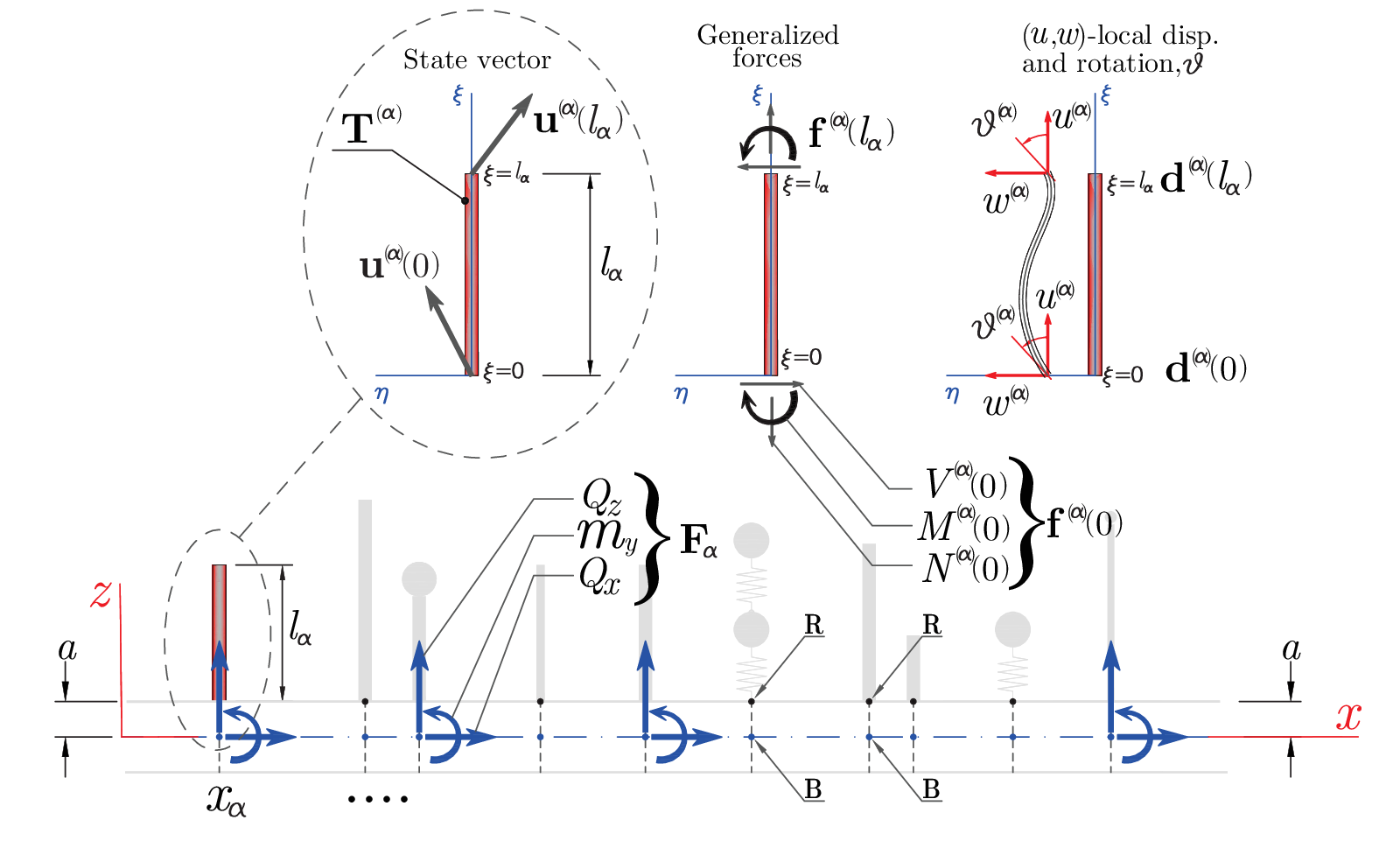} 
		\end{tabular}			
		\caption{Each resonator can be considered as a 1D waveguide itself along the local axis $\xi$, with  transfer matrix $\mathbf{T}^{(\alpha)}$ between attached-point to the beam ($\xi=0$) and the end-free-point ($\xi = l_\alpha$)}%
		\label{fig02}%
	\end{center}
\end{figure}

{
\begin{equation}
\begin{array}{lll}
	\text{Type of waves}				&			\text{Description}		&				\text{Equation}   \\ 	\hline  
	\text{Longitudinal waves}	  &			  \text{Constitutive	(normal)						 } 
	& 			\displaystyle	  \frac{\partial \mathcal{U}}{\partial x} = \frac{\mathcal{N}_x}{EA}	\\ \\
	&			 \text{Equilibrium $x$-axis			}
	&			 \displaystyle \frac{\partial \mathcal{N}_x}{\partial x} =\rho A \frac{\partial^2 \mathcal{U}}{\partial t^2} - q_x\\  \\
	\text{Flexural waves }			&			\text{Constitutive (shear)			 }
	& 			\displaystyle  \frac{\partial \mathcal{W}}{\partial x} = \frac{\mathcal{V}_z}{GA_z} + \Theta	\\  \\
	&			 \text{Equilibrium $z$-axis			}
	&			 \displaystyle \frac{\partial \mathcal{V}_z}{\partial x} =\rho A \frac{\partial^2 \mathcal{W}}{\partial t^2} - q_z\\  \\
	&			\text{Constitutive (bending)			 }
	& 			\displaystyle  \frac{\partial \Theta}{\partial x} = \frac{\mathcal{M}_y}{EI_y}		\\  \\
	&			 \text{Equilibrium $y$-axis			}
	&			 \displaystyle \frac{\partial \mathcal{M}_y}{\partial x} =\rho I \frac{\partial^2 \Theta}{\partial t^2} - \mathcal{V}_z- m_y \\ \\
		\hline 
\end{array}
\label{eq001}
\end{equation}
}
Above, $\mathcal{U}(x,t), \mathcal{W}(x,t)$ respectively denote the displacements in $x$ and $z$ direction of the reference axis and $\Theta(x,t)$ the cross section rotation. The functions $\mathcal{N}_x(x,t), \mathcal{V}_z(x,t), \mathcal{M}_y(x,t)$ represent the generalized forces associated to the previous kinematics variables, i.e. the normal and shear forces and bending moment at position $x$ and instant $t$ respectively. The geometrical properties are given by $A$ (area), $I_y$ (moment of inertia respect to $y$--axis) and the material properties are the density $\rho$, Young and shear modulus,  $E$ and $G$. The parameter $A_z$ has units of area and represents the shear reduced area, in general $A_z < A$. The magnitudes $q_x, \ q_z$ and $m_y$ represent the external forces and moment per unit of length respectively. These 6 variables can be rearranged in the following column state vector
\begin{equation}
\bm{\mathbf{U}}(x,t) = \{\mathcal{U}(x,t), \ \mathcal{W}(x,t) , \ \Theta(x,t), \ \mathcal{N}_x(x,t), \ \mathcal{V}_z(x,t), \ \mathcal{M}_y(x,t) \}^T.
\label{eq002}
\end{equation}
Assuming harmonic motion, we can write $\mathbf{U}(x,t) = \mathbf{u}(x) \, e^{i\omega t }$, with
\begin{equation}
	\mathbf{u}(x) = \{u(x), \ w(x), \ \theta(x), \ N(x), \ V(x), \ M(x)\}^T.
	\label{eq003}
\end{equation}
Along the paper, the reference central axes of the beam will be those shown in Fig.~\ref{fig01}, not being necessary to keep the subscript notation in variables of Eq.~\eqref{eq003}. The system of equations \eqref{eq001} is frequency--dependent and can be written in matrix form as
\begin{equation}
	\frac{\textrm{d} \mathbf{u}}{\textrm{d} x} = \mathbf{A} \, \mathbf{u} + \mathbf{q}(x),
	\label{eq004}
\end{equation}
where
\begin{equation}
	\mathbf{A} =
	\left[
	\begin{array}{cccccc}
	0  	& 0  & 0 & 1/EA & 0 & 0  \\ 
	0 & 0 & 1  & 0 &  1/GA_z & 0 \\ 
	0 & 0 & 0 & 0 & 0 & 1/EI_y  \\ 
   - \rho A \omega^2  &  0  & 0 & 	0 & 0 & 0  \\
	0 & - \rho A \omega^2 & 0 & 0 & 0 & 0 \\ 
	0 & 0 & - \rho I_y \omega^2 & 0 & -1 & 0 
	\end{array}
	\right]
	\ , \qquad 
	\mathbf{q}(x) = 
	\left\{
	\begin{array}{c}
		0  \\
		0 \\
		0 \\
		-q_x \\
		-q_z \\
		-m_y 
	\end{array}
\right\}
\label{eq004b}
\end{equation}
Notice that, under assumptions of homogeneous medium, longitudinal and flexural waves are decoupled. However, scatterers will introduce longitudinal forces induced by flexural motions of the beam, coupling the two types of waves. The term $ \mathbf{q}(x)$ represent the effect of external forces reduced to the $x$--axis and in turn it will be the superposition of two terms, say
\begin{equation}
	\mathbf{q}(x) = 	\mathbf{q}_s(x,\mathbf{u}) + 	\mathbf{q}_e(x)
	\label{eq004c} 
\end{equation}
where  $	\mathbf{q}_s(x,\mathbf{u})$ stands for the (unknown) forces applied on the $x$--axis due to the scattered field by the $N$ resonators and it depends on the value of the response at each point $\mathbf{u}(x_\alpha), \ 1 \leq \alpha \leq N$  
The remaining term $	\mathbf{q}_e(x)$ represents the known external loads on the waveguide which induces the ongoing incident field. According to Fig.
\ref{fig02} each resonator exerts on point $x=x_\alpha$ of the $x$--axis of the beam a force (two components) and a couple (moment), say $\{Q_x,Q_z,\mathfrak{M}_y\}$. We can then write by superposition
\begin{equation}
		\mathbf{q}_s(x,\mathbf{u}) = \sum_{\alpha = 1}^N 	\mathbf{q}_\alpha(x) 
		\label{eq118}
\end{equation}
where
\begin{equation}
	\mathbf{q}_\alpha(x)=
	- \left\{
	\begin{array}{c}
		0 \\ 0 \\ 0 \\ Q_x \\ Q_z \\ \mathfrak{M}_y 
	\end{array}
	\right\} \delta(x - x_\alpha)  \equiv \mathbf{Q}_\alpha \, \delta(x - x_\alpha) 
	\label{eq119}
\end{equation}
Plugging Eqs.\eqref{eq118} and \eqref{eq004c} into Eq. \eqref{eq004} we obtain the system of differential equations
\begin{equation}
	\frac{\textrm{d} \mathbf{u}}{\textrm{d} x} = \mathbf{A} \, \mathbf{u}  + \sum_{\alpha=1}^N \mathbf{Q}_\alpha \, \delta(x - x_\alpha) + \mathbf{q}_e(x)
	\label{eq005}
\end{equation}
The vectors $\mathbf{Q}_\alpha$ represent certain {\em source-terms} associated to the $\alpha$th resonator which in general depends linearly on the components of the state vector at $x=x_\alpha$. There exist then certain matrix $\mathbf{K}_\alpha$ associated to each scatterer, such that 
\begin{equation}
	\mathbf{Q}_\alpha = 	\mathbf{K}_\alpha \, \mathbf{u}(x_\alpha) \quad , \quad 1 \leq \alpha \leq N
		\label{eq006}
\end{equation}
where $\mathbf{u}(x_\alpha)$ stands for the state vector at the location of each scatterer and $\mathbf{K}_\alpha$ is a $6\times 6$ matrix which depends on the dynamical properties of the resonator. Our objective is double: (1) to propose a general procedure providing a closed--form for $\mathbf{K}_\alpha$ for any form of the attached beam-like resonator and (2) to derive an approximate solution under the assumption of {\em weak scattering} for 1D elastic waveguides. The derivations will provide precise quantification criteria for its application in terms of the properties of scatterers and their distribution along the beam. In this regards, we will define the conditions to apply the Born approximation in elastic 1D wavefields.\\

Let us determine the mechanical properties of the resonators, summarized in the matrix $\mathbf{K}_\alpha$. It is assumed that the scatterers are elastic structures any type as long as they can be modeled using the transfer matrix method along their length. Each resonator is linked to the main beam at point $x=x_\alpha, \ z = a$, where $a$ is the distance between the $x$--axis and the attachment point. We define a local $(\xi,\eta)$--axes associated to each resonator with length $l_\alpha$ (see Fig.~\ref{fig02}). Under this configuration, the scatterers will be excited under wave motion in the $xz$ symmetry plane, inducing vibrations in $(\xi,\eta)$-plane. Internal forces inside each resonator depend on the $\xi$-local coordinate and they are represented by $\mathbf{f}^{(\alpha)}(\xi) = \{N^{(\alpha)}(\xi),V^{(\alpha)}(\xi ),M^{(\alpha)}(\xi)\}^T$, say normal-force, shear-force and bending-moment, for any point $0 \leq \xi \leq l_\alpha$. At $\xi=l_\alpha$, 
it is $\mathbf{f}^{(\alpha)}(l_\alpha) = \mathbf{0}$, because the resonator has a free end, something that will be used later. On the other side, $\mathbf{f}^{(\alpha)}(0) = \{N^{(\alpha)}(0),V^{(\alpha)}(0),M^{(\alpha)}(0)\}^T$ represents the two forces and moment reactions on the resonator, as shown in Fig. \ref{fig02}. The forces $\mathbf{f}^{(\alpha)}(0)$, located at $x=x_\alpha$, $z=a$ are equivalent to the system $\mathbf{F}_\alpha = \{Q_x, \ Q_z, \ \mathfrak{M}_y\}^T$ at the $x$-axis, i.e. $z=0$. Thus, we have
\begin{equation}
	Q_x = - 	V^{(\alpha)}(0) \ , \qquad 
	Q_z =  	N^{(\alpha)}(0) \ , \qquad 
	\mathfrak{M}_y = a 	\, V^{(\alpha)}(0) 		+ 	M^{(\alpha)}(0) 
	\label{eq120}
\end{equation}
These relationships can be written in matrix form, yielding
\begin{equation}
	\mathbf{F}_\alpha =
	\left\{
	\begin{array}{c}
		Q_x \\
		Q_z \\
		\mathfrak{M}_y 
	\end{array}
	\right\} = 
	\left[
	\begin{array}{ccc}
	0  & -1  & 0    \\
	1  & 0  & 0    \\
	0  & a  & 1    
\end{array}
	\right]
		\left\{
	\begin{array}{c}
		N^{(\alpha)}(0) \\
		V^{(\alpha)}(0) \\
		M^{(\alpha)}(0)
	\end{array}
	\right\} \equiv \mathbf{H} \, \mathbf{f}^{(\alpha)}(0)
		\label{eq007}
\end{equation}
Compatibility relationships of displacements and rotations must be established between the attachment point on the beam--surface $\mathbf{R}(x=x_\alpha, z=a)$ and the point on the $x$--axis, say $\mathbf{B}(x=x_\alpha, z=0)$, see the location of points $\mathbf{R}$ and $\mathbf{B}$ in Fig. \ref{fig02}. The displacements and rotation of any point of the resonator along its length are arranged in the array $\mathbf{d}^{(\alpha)}(\xi) = \{u^{(\alpha)}(\xi),w^{(\alpha)}(\xi),\theta^{(\alpha)}(\xi)\}^T$ which represent longitudinal and transversal displacement and rotation refereed to the local axes $(\xi,\eta)$ (see Fig.~\ref{fig02}).  Displacements and rotation of the beam at point at $\mathbf{B}$ in $(x,z)$--axes are arranged in $\mathbf{d}(x_\alpha) = \{u(x_\alpha),w(x_\alpha),\theta(x_\alpha)\}^T$. Invoking the kinematics of the main-beam cross section in terms of $\mathbf{d}(x)$ and taking into account the different reference directions of each variable we have that
\begin{equation}
		u^{(\alpha)}(0)  = w(x_\alpha) \ , \qquad 
    	w^{(\alpha)}(0)=  - u(x_\alpha) + a \, \theta (x_\alpha) \  , \qquad 
	\theta^{(\alpha)}(0) = \theta(x_\alpha)
	\label{eq121}
\end{equation}
which again can be written in matrix form as
\begin{equation}
	\mathbf{d}^{(\alpha)}(0) =
	\left\{
	\begin{array}{c}
		u^{(\alpha)}(0) \\
		w^{(\alpha)}(0) \\
		\theta^{(\alpha)}(0)
	\end{array}
	\right\} = 
	\left[
	\begin{array}{ccc}
		0  & 1  & 0    \\
		-1  & 0  & a    \\
		0  & 0  & 1    
	\end{array}
	\right]
	\left\{
	\begin{array}{c}
		u(x_\alpha) \\
		w(x_\alpha) \\
		\theta(x_\alpha)
	\end{array}
	\right\} \equiv \mathbf{H}^T \, \mathbf{d}(x_\alpha)
		\label{eq008}
\end{equation}
The forces at point $\mathbf{B}$ induced by the resonator are a consequence of the vibrations produced by the motion in the beam and transmitted along the body of the resonator. This interaction is captured in a relationship between forces $\mathbf{F}_\alpha$  and displacements  $\mathbf{d}(x_\alpha)$. Based on the one-dimensional definition of the resonator along axis $\xi$ for $0\leq \xi \leq l_\alpha$, the transfer matrix method can be used to relate the displacements--forces state--vector  $\mathbf{u}^{(\alpha)}(\xi) = \{\mathbf{d}^{(\alpha)}(\xi),\mathbf{f}^{(\alpha)}(\xi)\}^T$ at locations $\xi=0$ and $\xi = l_\alpha$. Thus, denoting by $\mathbf{T}^{(\alpha)}$ to the transfer matrix between both stations $\xi=0$ and $\xi=l_\alpha$, it yields
\begin{equation}
	\mathbf{u}^{(\alpha)}(l_\alpha) = 
	\left\{
	\begin{array}{c}
		\mathbf{d}^{(\alpha)}(l_\alpha)	\\
		\mathbf{f}^{(\alpha)}(l_\alpha)	
	\end{array}
	\right\} = 
			\mathbf{T}^{(\alpha)}	\\
	\left\{
\begin{array}{c}
	\mathbf{d}^{(\alpha)}(0)	\\
	\mathbf{f}^{(\alpha)}(0)	
\end{array}
\right\} \equiv
			\mathbf{T}^{(\alpha)}	
			\mathbf{u}^{(\alpha)}(0)	
\label{eq009}
\end{equation}
In the current derivations, the transfer matrix $\mathbf{T}^{(\alpha)}$ is a $6 \times 6$ array which depends on the resonator mechanical properties and on frequency, including possible added elastic or mass devices along the length $l_\alpha$, resulting in such case  the product of the different transfer matrices associated to each segment \cite{Rui-2019}. Splitting up the matrix $\mathbf{T}^{(\alpha)}$ in four $3 \times 3$ block-matrices, we can relate both displacements and internal forces at both ends, $\xi=0,l_\alpha$, resulting in
\begin{equation}
\left\{
	\begin{array}{c}
	\mathbf{d}^{(\alpha)}(l_\alpha)	\\
	\mathbf{f}^{(\alpha)}(l_\alpha)	
\end{array}
\right\} = 
 \left[
 \begin{array}{cc}
    \mathbf{T}^{(\alpha)}_{dd} & \mathbf{T}^{(\alpha)}_{df} \\
   \mathbf{T}^{(\alpha)}_{fd} & \mathbf{T}^{(\alpha)}_{ff} 
 \end{array} 
 \right]
\left\{
\begin{array}{c}
	\mathbf{d}^{(\alpha)}(0)	\\
	\mathbf{f}^{(\alpha)}(0)	
\end{array}
\right\} 
\label{eq010}
\end{equation}
The opposite edge of the resonator at $\xi=l_\alpha$ is forces--free, therefore $	\mathbf{f}^{(\alpha)}(l_\alpha) = \mathbf{0}$ leading to the equation
\begin{equation}
    \mathbf{0} = \mathbf{T}^{(\alpha)}_{fd} \mathbf{d}^{(\alpha)}(0) +  \mathbf{T}^{(\alpha)}_{ff} \mathbf{f}^{(\alpha)}(0)
    \label{eq011}
\end{equation}
yielding
\begin{equation}
	 \mathbf{f}^{(\alpha)}(0) = - \left[  \mathbf{T}^{(\alpha)}_{ff}\right]^{-1} \, \mathbf{T}^{(\alpha)}_{fd} \,  \mathbf{d}^{(\alpha)}(0) 
	\label{eq012}
\end{equation}
From Eqs. ~\eqref{eq007}, \eqref{eq012} and \eqref{eq008}, we obtain the following relationship between beam forces $\mathbf{F}_\alpha$ and displacements $\mathbf{d}(x_\alpha)$ in global axes $(x,z)$
\begin{equation}
	\mathbf{F}_\alpha  = - \mathbf{H} \, \left.  \mathbf{T}^{(\alpha)}_{ff} \right. ^{-1} \, \mathbf{T}^{(\alpha)}_{fd} \, \mathbf{H}^T \, \mathbf{d}(x_\alpha)
	\label{eq013}
\end{equation}
The point-forces can then be introduced into the differential equation \eqref{eq004} using the Dirac-delta function, resulting
\begin{equation}
	\mathbf{q}_\alpha(x)=
	- \left\{
	\begin{array}{c}
		\mathbf{0}	\\
		\mathbf{F}_{\alpha}
	\end{array}
	\right\} \delta(x - x_\alpha) 
	=
	\left[
	\begin{array}{cc}
		\mathbf{0}  & \mathbf{0} \\
	 \mathbf{H} \, \left.  \mathbf{T}^{(\alpha)}_{ff}  \right. ^{-1} \, \mathbf{T}^{(\alpha)}_{fd} \, \mathbf{H}^T  & \mathbf{0} 
	\end{array} 
	\right]
	\left\{
	\begin{array}{c}
		\mathbf{d}(x_\alpha)	\\
		\mathbf{f}(x_\alpha)
	\end{array}
	\right\} \delta(x - x_\alpha) \equiv \mathbf{K}_\alpha \, \mathbf{u}(x_\alpha)\, \delta(x-x_\alpha)
	\label{eq014}
\end{equation}
where the matrix $\mathbf{K}_\alpha$ is defined as
\begin{equation}
	\mathbf{K}_\alpha = 
	\left[
\begin{array}{c|c}
	\mathbf{0}  & \mathbf{0} \\
	\hline
	\mathbf{H} \, \left.  \mathbf{T}^{(\alpha)}_{ff}\right. ^{-1} \, \mathbf{T}^{(\alpha)}_{fd} \, \mathbf{H}^T  & \mathbf{0} 
\end{array} 
\right]	
	\label{eq015}
\end{equation}
then we can write $\mathbf{Q}_\alpha = \mathbf{K}_\alpha \, \mathbf{u}(x_\alpha)$ as the pointwise forces and induced by the incident wave. Within the matrix $\mathbf{K}_\alpha$ we can find all properties of the resonators, including the set of resonances which will correspond to the roots of the equation 
\begin{equation}
	\det \left[\mathbf{T}^{(\alpha)}_{ff} \right]= 0 \qquad \to \qquad \text{Resonance frequencies: } \ \omega_1,\omega_2, \cdots
	\label{eq015b}
\end{equation}

\section{Solution of the multiple scattering problem}
%


The closed-form expression found for the matrix $ \mathbf{K}_\alpha $ enables to write Eq.~\eqref{eq004} in terms of the radiating wave outgoing from the resonators as
\begin{equation}
		\frac{\textrm{d} \mathbf{u}}{\textrm{d} x} = \mathbf{A} \, \mathbf{u} + 
	\sum_{\alpha+1}^{N} \mathbf{K}_\alpha \mathbf{u}(x_\alpha) \, \delta(x - x_\alpha) 	+ \mathbf{q}_e(x) 
	\label{eq036}
\end{equation}
where $\mathbf{q}_e(x)$ represents the vector of external forces. The total solution can be written as 
\begin{equation}
	\mathbf{u}(x) = \bm{\psi}_0(x) + 	\mathbf{u}_s(x)
	\label{eq037}
\end{equation}
where $\bm{\psi}_0(x)$ denotes the incidence wavefield generated by the external forces and is solution of the problem
\begin{equation}
	\frac{\textrm{d} \bm{\psi}_0}{\textrm{d} x} = \mathbf{A} \, \bm{\psi}_0 + \mathbf{q}_e(x) 
	\label{eq036a}
\end{equation}
and $\mathbf{u}_s(x)$ is the scattered field, which is  solution of 
\begin{equation}
	\frac{\textrm{d} \mathbf{u}_s}{\textrm{d} x} = \mathbf{A} \, \mathbf{u}_s + 
	\sum_{\alpha+1}^{N} \mathbf{K}_\alpha \mathbf{u}(x_\alpha) \, \delta(x - x_\alpha) 	
	\label{eq038}
\end{equation}
The solution will be presented in terms of the Green matrix of the system, defined as
\begin{equation}
	\left( \frac{\textrm{d}}{\textrm{d} x} -  \mathbf{A} \right) \mathbf{G} = \mathbf{I} \, \delta (x)
	\label{eq039}
\end{equation}
In \ref{ap_green} the analytical form of this matrix is derived in terms of eigenvalues and eigenvectors of matrix $\mathbf{A}$. The general solution of Eq.~\eqref{eq036} can then be written as
\begin{equation}
	\mathbf{u}(x) = \bm{\psi}_{0}(x) + \mathbf{u}_s(x) = \bm{\psi}_{0}(x) + \sum_{\beta=1}^N \mathbf{G}(x-x_\beta) \, \mathbf{K}_\beta \, \mathbf{u}(x_\beta)
	\label{eq040}
\end{equation} 
After evaluating Eq.~\eqref{eq040} at $x_1, \ldots, x_N$, the following system of linear equation arises
\begin{equation}
	\mathbf{u}(x_\alpha) - \sum_{\beta=1}^N \mathbf{G}(x_\alpha-x_\beta) \, \mathbf{K}_\beta \, \mathbf{u}(x_\beta)= \bm{\psi}_{\text{0}}(x_\alpha)
	\quad , \quad \ 1 \leq \alpha \leq N
	\label{eq041} 
\end{equation}
Let us now introduce the following auxiliary variables associated to each resonator
\begin{equation}
	\bm{\phi}(x_\alpha) = \left[ \mathbf{I} - \mathbf{G}(0^+) \mathbf{K}_\alpha\right] \, \mathbf{u}(x_\alpha)
	\label{eq042}
\end{equation}
After some rearrangements, the system of equations \eqref{eq041} can be written in terms of the new variables defined above as
\begin{equation}
	\left[
	\begin{array}{cccc}
		\mathbf{I}			&			- \mathbf{G}(x_1-x_2) \mathbf{t}_2   & \cdots 	&- \mathbf{G}(x_1-x_N) \mathbf{t}_N \\
		- \mathbf{G}(x_2-x_1) \mathbf{t}_1  &  \mathbf{I}			   & \cdots 	&- \mathbf{G}(x_2-x_N) \mathbf{t}_N \\
		\vdots					 & 				\vdots															 &   \ddots  &   \vdots  \\
		- \mathbf{G}(x_N-x_1) \mathbf{t}_1			&	- \mathbf{G}(x_N-x_2) \mathbf{t}_2   & \cdots 	& 	\mathbf{I}		 
	\end{array}
	\right]
	\left\{
	\begin{array}{c}
		\bm{\phi}(x_1) \\ 		\bm{\phi}(x_2) \\ \vdots \\ 		\bm{\phi}(x_N)
	\end{array}\right\}	
	=
	\left\{
	\begin{array}{c}
		\bm{\psi}_0(x_1) \\ 		\bm{\psi}_0(x_2) \\ \vdots \\ 		\bm{\psi}_0(x_N)
	\end{array}\right\}	
	\label{eq043}
\end{equation} 
where the new matrices $\mathbf{t}_\alpha$ are defined as
\begin{equation}
	\mathbf{t}_\alpha = \mathbf{K}_\alpha \, \left[ \mathbf{I} - \mathbf{G}(0^+) \mathbf{K}_\alpha\right] ^{-1}
	\label{eq044}
\end{equation}
Eq. \eqref{eq043} can be written in compact form as
\begin{equation}
	\left[\mathbf{I} - \mathbf{g} \right] \, \mathbf{\Phi} = \mathbf{\Psi}_{\text{0}}
	\label{eq045}
\end{equation}
where the matrix $ \mathbf{g} = \hat{\mathbf{G}} \, \hat{\mathbf{T}} $ and
\begin{equation}
	\hat{\mathbf{G}} = 
	\left[
	\begin{array}{cccc}
		\mathbf{0}			&			\mathbf{G}(x_1-x_2)   & \cdots 	& \mathbf{G}(x_1-x_N) \\
		\mathbf{G}(x_2-x_1)   &  \mathbf{0}			   & \cdots 	& \mathbf{G}(x_2-x_N)  \\
		\vdots					 & 				\vdots															 &   \ddots  &   \vdots  \\
		\mathbf{G}(x_N-x_1)			&	 \mathbf{G}(x_N-x_2)   & \cdots 	& 	\mathbf{0}		 
	\end{array}
	\right] \ , \quad
	\hat{\mathbf{T}}  = 
	\left[
	\begin{array}{cccc}
		\mathbf{ \mathbf{t}_1}			&			\cdots 	&  \mathbf{0} \\
		\vdots					 & 				 \ddots  &   \vdots  \\
		\mathbf{0}			&	\cdots 	& 	\mathbf{t}_N		 
	\end{array}
	\right]
	\label{eq046}
\end{equation}
The scattering coefficients $\bm{\Phi}$ emerge from the solution of the above equation which in terms of the inverse matrix has the form
\begin{equation}
	\mathbf{\Phi} = \left[\mathbf{I} - \mathbf{g} \right]^{-1} \mathbf{\Psi}_{\text{0}}
	\label{eq047}
\end{equation}
And the total wave field can be obtain from 
\begin{equation}
	\mathbf{u}(x) = \bm{\mathbf{\psi}}_{\text{0}}(x)  + \sum_{\alpha=1}^N \mathbf{G}(x-x_\alpha) \,  \mathbf{t}_\alpha \, \bm{\phi}(x_\alpha)
	\label{eq048}
\end{equation}
The solution of Eq.~\eqref{eq047} and the wavefield of Eq. \eqref{eq048} provide the exact approach to the problem for any frequency.

\subsection{General conditions for weak scattering and the Born approximation in beams}

As shown above, the scattering coefficients can be found from the solution of Eq.~\eqref{eq047}. If the elements of the matrix $\mathbf{g}$ are sufficiently small, the inverse can be efficiently evaluated  by means of the Neumann series expansion, namely
\begin{equation}
	\left[\mathbf{I} - \mathbf{g} \right]^{-1} = \mathbf{I} + \mathbf{g} + \mathbf{g}^2 + \cdots = \sum_{n=0}^\infty \mathbf{g}^n
	\label{eq086}
\end{equation}
%
%
The necessary and sufficient condition to guarantee that such an expansion can be carried out is intimately linked to the magnitude of the spectrum (set of eigenvalues) of the matrix $\mathbf{g}$. In fact, if the spectral radius $\rho(\mathbf{g})$ is less than unity, we can ensure that the series converges. Moreover, the number of terms needed to reach a certain truncation error increases as the magnitude of $\rho(\mathbf{g})$ approaches the unity, provided that $\rho(\mathbf{g})<1$ \cite{Householder-1964,Wilkinson-1988}. The scattering induced by the $N$ resonators is then said to be weak. On the other hand, there are situations, especially around scatterer resonances, where $\rho(\mathbf{g})>1$ holds and consequently the inverse cannot be approximated by the expansion of Eq. \eqref{eq086}. Such situations are associated with strong scattering.

Let us see the consequences of the inequality $\rho(\mathbf{g}) < 1$ in our problem for a given frequency. As shown in Eq. \eqref{eq046}, the terms of matrix $\mathbf{g} = \hat{\mathbf{G}} \hat{\mathbf{T}}$  are formed by the product of two matrices. Matrix $\hat{\mathbf{G}}$ represents the coupling of the different propagating modes scattered by the resonator array. On the other, the matrix $\hat{\mathbf{T}}$ is a diagonal-blocks matrix the resonators impedances. Moreover, both matrices are frequency dependent and it is therefore expected that the value of $\mathbf{g}$ will increase around the resonances of the system. Additionally, it will also affect how these ones are arranged, allowing the appearance of Bragg peaks due to induced periodicities, something that is reflected in the $\mathbf{G}$ matrix. \\

Assuming then that $\rho(\mathbf{g}) < 1$,  the solution can be obtained recursively defining the sequence
\begin{equation}
	\bm{\Phi}_n = \mathbf{g} \,  	\bm{\Phi}_{n-1} \ , \qquad   \bm{\Phi}_0 =  \mathbf{\Psi}_{\text{0}}
	\label{eq087}
\end{equation}
Thus, the scattering coefficients are the result of the series
\begin{equation}
	\mathbf{\Phi} = \left[\mathbf{I} - \mathbf{g} \right]^{-1} \mathbf{\Psi}_{\text{0}} =
	\left(  \mathbf{I} + \mathbf{g} + \mathbf{g}^2 + \cdots \right) \mathbf{\Psi}_{\text{0}} =
	\mathbf{\Phi}_0 + \mathbf{\Phi}_1 + \mathbf{\Phi}_2 + \cdots	= \sum_{n=0}^\infty \bm{\Phi}_n 
	\label{eq088}
\end{equation}
The above expression captures the essence of the multiple scattering since the effect of  higher order scattered waves are associated to higher order terms of the sequence. The zero-order solution $	\mathbf{\Phi} \approx 	\mathbf{\Phi}_0$ corresponds to consider 
\begin{equation}
	\bm{\phi}(x_\alpha) \approx \bm{\psi}_0(x_\alpha)
	\label{eq0106}
\end{equation}
The wavefield associated is then straightforward and can be written as
\begin{equation}
	\mathbf{u}(x) \approx \bm{\mathbf{\psi}}_{\text{0}}(x)  + \sum_{\alpha} \mathbf{G}(x-x_\alpha) \,  \mathbf{t}_\alpha \, \bm{\psi}_0(x_\alpha)
	\label{eq108}
\end{equation}
Higher order scattering waves are neglected in this approximation. As an extension of the concept used in other fields of wave physics, the so-obtained wavefield  is called {\em Born--approximation} in the context of  1D elastic waveguides. As Eq. \eqref{eq108} shows, the scattered wavefield of Born-aproximation is proportional to the intensity of the scatterers and also depends on their  relative positions via the Green matrix. Considering higher-order scattering is equivalent to add new terms to the series $\mathbf{\Phi} = \mathbf{\Phi}_0 + \mathbf{\Phi}_1 + \mathbf{\Phi}_2 + \cdots$. Now the following question arises: Can we quantify mathematically the quality of Born--approximation? Indeed, the accuracy is closely related to the magnitude of the spectral radius $\rho(\mathbf{g})$, which is a closed-form dimensionless measurement of the scattering intensity. Thus, lower values of  $\rho(\mathbf{g})$ will be associated with good agreement of the response approximated by Eq.\eqref{eq108} as will be validated in the numerical examples. In this analysis we will go further and in the next point an approximation of the spectral radius $\rho(\mathbf{g})$ is proposed. We know that the spectral radius corresponds to the maximum eigenvalue in magnitude. The analysis carried out will allow us to dissociate two very important characteristics of scattered media: (i) the relative position between scatters and (ii) the pointwise scattering intensity of each resonator. \\

\subsection{Approximate solution of the spectral radius $\rho(\mathbf{g})$}

As stated above, the general solution of the scattered field require the computation of the scattering coefficients given by
$\bm{\Phi} = \left[\mathbf{I} - 	\mathbf{g} \right]^{-1} \, \bm{\Psi_\text{0}}$. The weak scattering approach can be assumed provided that the convergence of the Neuman series \eqref{eq088} is ensured, which in turn occurs if the spectral radius of the matrix $\mathbf{g} = \hat{\mathbf{G}} \hat{\mathbf{T}}$ is less than unity, i.e. $\rho (\hat{\mathbf{G}} \hat{\mathbf{T}})<1$. Although it is always possible to compute the eigenvalues of the matrix and evaluate the spectral radius, it may be useful from a practical point of view to have approximate expressions that allow predicting its value a priori. This would also allow us to go deeper into the physical insight of weak scattering since such an expression, as we will see below, contains explicitly all the parameters of the problem. \\

\begin{figure}[h]%
	\begin{center}
		\includegraphics[width=10cm]{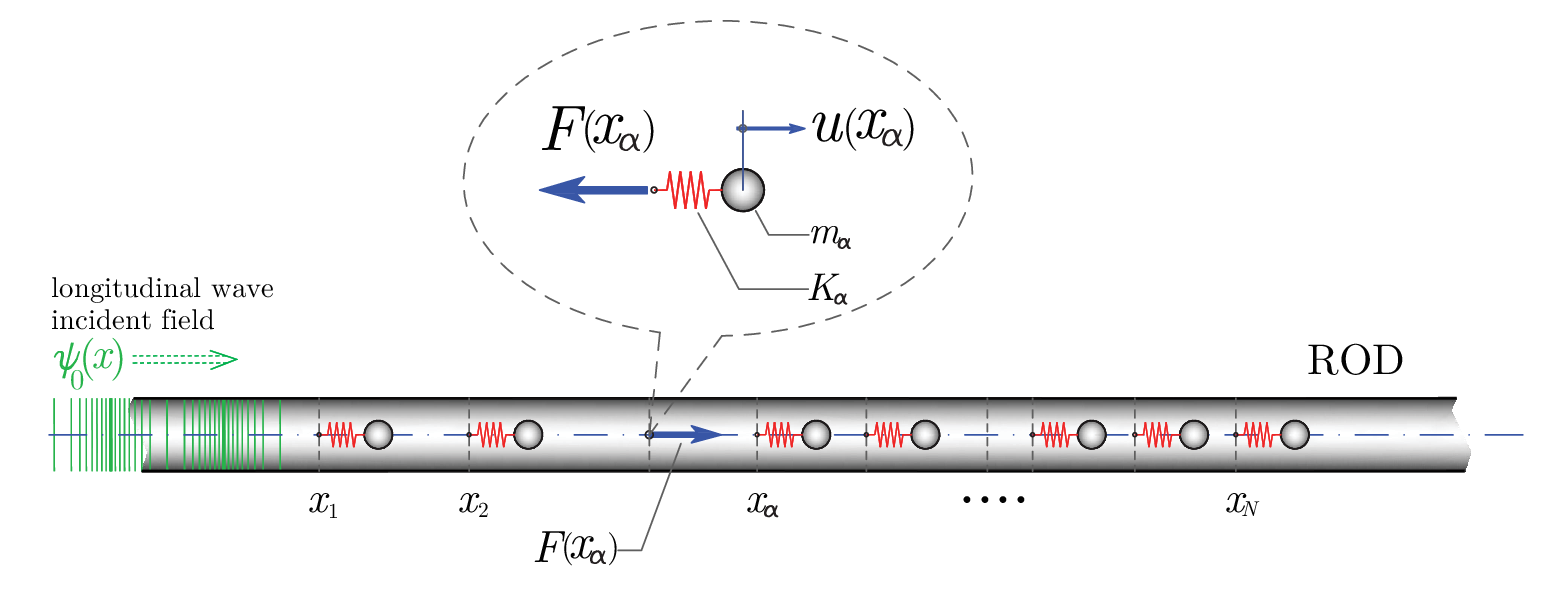} \\	
		\caption{Longitudinal wave propagation along a rod with an arrangement of $N$ single-resonance scatterers.}%
		\label{fig04}%
	\end{center}
\end{figure}

Our goal is to investigate deeper the structure of scattering matrix $\mathbf{g}$ decoupling the different effects: on one side the terms which depend only on the relative positions between scatterers  and on the other side, on the relative impedance of the resonators. To that end, a simpler rod with $N$ pointwise single resonators will be addressed, see Fig. \ref{fig04}. In this case, only longitudinal waves are under consideration and it is not necessary to write the wavefield in vector form $\mathbf{u}(x)$ since we can establish the scattering relationships in terms of  horizontal displacements. Thus, at $x = x_\alpha$, the spring-force and the rod displacement are linked by
\begin{equation}
	F(x_\alpha) = - \frac{K_{\alpha} \, \omega^2}{\omega^2 - \omega_{\alpha}^2}  u(x_\alpha) \equiv  \hat{K}_{\alpha}\, u(x_\alpha) \quad
	\label{eq089}
\end{equation}
where $K_\alpha = m_\alpha \, \omega_{\alpha}^2$ are the resonator parameters. Assuming now the displacements incidence  field is $\psi_{\text{0}}(x)$, then the total field, measured as the sum of incidence and scattered field can be written as
\begin{equation}
	u(x) = \psi_{\text{0}}(x) + \sum_{\alpha=1}^{N}  G(x-x_\alpha) \,  \hat{K}_\alpha  \, u(x_\alpha)
	\label{eq090}
\end{equation}
where $G(x)$ stands for the Green function relating forces with displacements (with units of flexibility, i.e. m/N). It corresponds to the $(1,4)$--entree of matrix $\mathbf{G}(x)$, which after some algebra can be expressed as
\begin{equation}
	G(x) =  \frac{1}{2 i k EA } e^{-i k \left|x\right|}
	\label{eq091}
\end{equation}
where $k = \omega \sqrt{\rho A / EA}$ is the wavenumber of longitudinal waves. Eq.~\eqref{eq090} represents a scalar version of Eq.~\eqref{eq040}, therefore the response on the beam can be found after evaluating Eq.~\eqref{eq090} at each $x_\nu, \quad 1 \leq \nu \leq N$ leading to the following system of $N$ equations
\begin{equation}
	u(x_\alpha) - \sum_{\beta=1}^{N}  G(x_\alpha-x_\beta) \, \hat{K}_\beta \, u(x_\beta) =  \psi_{\text{0}}(x_\alpha) 
	\ , \quad 1 \leq \alpha \leq N
	\label{eq092}
\end{equation}
Rearranging in matrix form
\begin{equation}
	\begin{bmatrix}
		1 - G(0)\hat{K}_1 			&  -G(x_1-x_2)\hat{K}_2 & \cdots & - G(x_1-x_N)\hat{K}_N   \\
		-G(x_2-x_1)\hat{K}_1 	&  1 - G(0)\hat{K}_2 & \cdots & - G(x_2-x_N)\hat{K}_N       \\		
		\vdots								 &  \vdots					& \ddots	&	\vdots						      \\
		-G(x_N-x_1)\hat{K}_1 	&  -G(x_N-x_2)\hat{K}_2 & \cdots & 1 - G(0)\hat{K}_N 
	\end{bmatrix}
	\begin{Bmatrix}
		u(x_1) \\
	    u(x_2) \\
		\vdots  \\
		u(x_N) 
	\end{Bmatrix}
	=
	\begin{Bmatrix}
		\psi_0(x_1) \\
		\psi_0(x_2) \\
		\vdots  \\
		\psi_0(x_N) 
	\end{Bmatrix}
	\label{eq093}
\end{equation}
The Green function can be written in the form
\begin{equation}
	G(x) = G(0) \, e^{-i k \left|  x \right|} 
	\label{eq096}
\end{equation}
The reason to adopt this new expression is justified later, when addressing the general case considering both longitudinal and flexural propagating waves. Let us introduce now the following auxiliary variables 
\begin{equation}
	\phi(x_\alpha)  = \left[1 - G(0) \hat{K}_\alpha\right] \, u(x_\alpha)  \quad , \quad \tau_\alpha = \frac{G(0) \, \hat{K}_\alpha}{1 - G(0) \hat{K}_\alpha}	
	\label{eq097}
\end{equation}
Then the system of equations can be written in the form
\begin{equation}
	\begin{bmatrix}
		1		&  - \tau_2 e^{-i k \left|x_1-x_2 \right|} & - \tau_3  e^{-i k \left|x_1-x_3 \right|} & \cdots & - \tau_N e^{-i k \left|x_1-x_N \right|} \\
		-\tau_1  e^{-i k \left|x_2-x_1 \right|} & 1	&   - \tau_3  e^{-i k \left|x_2-x_3 \right|}& \cdots & - \tau_N  e^{-i k \left|x_2-x_N \right|} \\ 
		\vdots											&  			\vdots  									&  \vdots & \ddots & \vdots  \\
		- \tau_1  e^{-i k \left|x_N-x_1 \right|} & 	 - \tau_2  e^{-i k \left|x_N-x_2 \right|}		&   - \tau_3  e^{k_2 \left|x_N-x_3 \right|}& \cdots & 1 \\ 
	\end{bmatrix}
	\begin{Bmatrix}
		\phi(x_1) \\
		\phi(x_2) \\
		\vdots  \\
		\phi(x_N) 
	\end{Bmatrix}
	=
	\begin{Bmatrix}
		\psi_0(x_1) \\
		\psi_0(x_2) \\
		\vdots  \\
		\psi_0(x_N) 
	\end{Bmatrix}
	\label{eq098}
\end{equation}
or in matrix form
\begin{equation}
	\left[\mathbf{I}_N - \bm{\mathcal{G}}(k) \, \bm{\tau}  \right] \, 	\mathbf{\Phi} = \mathbf{\Psi}_{\text{0}}
	\label{eq099}
\end{equation}
where $\bm{\mathcal{G}}(k)$ is the matrix formed by exponential functions evaluated at the relative positions of the scatterers and asociated to the propagating wavenumber $k$. 
\begin{equation}
	\bm{\mathcal{G}}(k) =
	\begin{bmatrix}
		0													&  e^{-ik \left|x_1-x_2 \right|} & e^{-ik \left|x_1-x_3 \right|} & \cdots & e^{-ik \left|x_1-x_N \right|} \\
		e^{-ik\left|x_2-x_1 \right|} & 0 														&   e^{-ik \left|x_2-x_3 \right|}& \cdots & e^{-ik\left|x_2-x_N \right|} \\ 
		\vdots											&  			\vdots  									&  \vdots & \ddots & \vdots  \\
		e^{-ik \left|x_N-x_1 \right|} & 	 e^{-ik \left|x_N-x_2 \right|}		&   e^{-ik \left|x_N-x_3 \right|}& \cdots & 0 \\ 
	\end{bmatrix}
	\ , \qquad
	\bm{\tau} = 
	\begin{bmatrix}
		\tau_1			&  \cdots & 0 \\
		\vdots			& \ddots	&	\vdots	\\
		0					&  \cdots & \tau_N
	\end{bmatrix}
	\label{eq100}
\end{equation}
Note that somehow the scattering coefficients are separated in terms of the form $e^{-ik \left|x_\alpha-x_\beta \right|}$ which mainly depend on the relative positions of the scatteres respect to the propagating wavelength $2\pi/k$, and $\tau_\alpha$ which depends on the contrast of elastic impedance between  resonator and rod. The solution can be expressed as
\begin{equation}
	\bm{\Phi} = \left[\mathbf{I}_N - 	\bm{\mathcal{G}}(k) \bm{\tau} \right]^{-1} \, \bm{\Psi_\text{0}}
	\label{eq101}
\end{equation}
If all scatterers have the same properties $\tau_1 = \cdots = \tau_N$ , then we can write
\begin{equation}
	\rho(\mathbf{g}) = \rho \left[\bm{\mathcal{G}}(k)\right] \cdot  \left|\tau_\alpha \right|
	\label{eq103}
\end{equation}
In  general, with different scatterers we can conjecture that the following approximation will approach accurately enough for our propose of finding  the spectral radius of $\mathbf{g}$.
 \begin{equation}
 	\rho(\mathbf{g}) \approx \rho \left[\bm{\mathcal{G}}(k)\right]  \cdot \max_{\alpha } \left|\tau_\alpha\right| = 
 	\rho \left[\bm{\mathcal{G}}(k)\right]  \cdot \rho(\bm{\tau})
 	\label{eq104}
 \end{equation}
 The result of this proposed approximation will be validated numerically later. In Eq. \eqref{eq104} we identify separately the effect of scatterers relative positions respect to the traveling wavelength $2\pi / k$ and, on the other side, a dimensionless measurement of the relative impedance between each scatterer $\hat{K}_\alpha$ (N/m) and the medium $G(0)$, measured in terms of flexibility (m/N). As shown in Eq.\eqref{eq097}, the parameter $\tau_\alpha$ is dimensionless and in the particular case of longitudinal waves takes the form
 \begin{equation}
 	\tau_\alpha = \frac{1}{1 - \left(1 - \frac{\omega_\alpha^2}{\omega^2}\right) \frac{2 i k EA}{\hat{K}_\alpha}}
 	\label{eq104b} 
 \end{equation}
Physically, the reflection and transmission coefficients of a single resonator ($N=1$) are written in terms of this parameter as
\begin{equation}
	R = \tau_\alpha e^{-2 i k x_\alpha} \quad , \quad T = 1 + \tau_\alpha \qquad , \quad (\textnormal{for } \  N=1)
\end{equation}
 Therefore, the elements of the $\bm{\tau}$ matrix represent the intensities of the scatterers as isolated. The joint effect of the relative positions of the scatterers with respect to the wavelength is given by the matrix $\bm{\mathcal{G}}(k) $. Using the properties of the spectral radius in relation to the matrix norm we can establish an upper bound of $	\rho(\mathbf{g})$.  It is known that the spectral radius of a matrix is bounded by any consistent norm. Thus,  we have 
\begin{equation}
	\rho(\mathbf{g})   \leq \left\| 	\bm{\mathcal{G}}(k)   	  \, \bm{\tau} \right\|
	\leq \left\| \bm{\mathcal{G}}(k) \right \|  \cdot	\left\| \bm{\tau} \right\| = \left\| \bm{\mathcal{G}}(k) \right \|  \cdot	\rho(\bm{\tau})
	\label{eq080}
\end{equation}
where the last equality holds because $\bm{\tau} $ is a diagonal matrix. For both norms $ \left\| \bullet \right \|_1$ and $ \left\| \bullet \right \|_\infty$ it follows that
\begin{equation}
	\left\| \bm{\mathcal{G}}(k) \right \|_1 =  	\left\| \bm{\mathcal{G}}(k) \right \|_\infty =  N-1 \
	\label{eq081}
\end{equation}
Therefore, we can ensure that for any frequency we have
\begin{equation}
	\rho(\mathbf{g}) \leq (N-1) \cdot \rho(\bm{\tau})
	\label{eq082}
\end{equation}
Therefore, the above inequality allow us to propose a condition for weak scattering in simple terms of the number of scatterers $N$ and the properties of each scatterer. Thus, we can assure that the Neumann series which define the weak scattering approximation converges provided that $(N-1) \cdot  \max_{\alpha }  \left| \tau_\alpha \right| < 1$. 
A quick inspection of this condition highlights the two main parameters of the scattering problem: the number of scatters and the intensity of each one independently. However, the consideration of the upper bound $\rho [\bm{\mathcal{G}}(k)]  \leq N-1$ turns out to be very conservative since does not take into account the interaction between the positions of the different scatters respect to the wavelength. In fact, the equality holds only for periodic distributions of scatterers at frequencies corresponding to Bragg peaks. In such case, every exponential term reduces to the unity and the maximum eigenvalue is indeed $N-1$. Let us see that  a more detailed expression for the estimation of the spectral radius 
$\rho [\bm{\mathcal{G}}(k)]$ can be obtained using the well-known power iteration method.\\

The spectral radius of matrix $\bm{\mathcal{G}}(k)$, given in Eq. \eqref{eq100} and  formed by exponential terms associated to wavenumber $k$ is not available analytically but play an important role in the scattering problem. Our goal is to obtain a reasonable estimation using the algorithm for computation of the largest eigenvalue: the power iteration method. In particular, we will make use just only of the first iteration so that analytical expressions can be derived.  The following derivations are valid for a set of $N$ scatters distributed on positions $x_1,\ldots,x_N$, where $x_1 < x_2 < \cdots < x_N$. Consider a propagating wave of any nature (longitudinal, bending or shear) with wavelength $2 \pi / k$ is traveling along the beam. The wavenumber $k$ is one of the possible propagating modes admissible in the beam/rod. Let us consider the $N$--dimensional and unitary complex vector as the first guess in the power-iteration method
\begin{equation}
	\mathbf{y}_0 = \{e^{i k  x_1},\ldots,e^{i k  x_N}\}/ \sqrt{N}
	\label{eq083}	
\end{equation}
which verifies that $\mathbf{y}_0^H \mathbf{y}_0 =1$, where $(\bullet)^H$ denotes conjugate-transponse operator. After some algebra, the first iteration of the power method leads to the expression
\begin{equation}
	\rho \left[\bm{\mathcal{G}}(k) \right] \approx \left| \frac{\mathbf{y}_0^H \bm{\mathcal{G}}(k) \, \mathbf{y}_0 }{\mathbf{y}_0^H \mathbf{y}_0} \right|
	= \left| \frac{N-1}{2} + \frac{1}{N}\sum_{\alpha=1}^{N} \sum_{\beta < \alpha} e^{-2i k (x_\alpha - x_\beta)}  \right|
	\label{eq084}
\end{equation}
Thus based on this expression, we propose the following estimation of the spectral radius for practical proposes in this particular case of rods with resonators
\begin{equation}
	\rho(\mathbf{g}) \approx
	 \left| \frac{N-1}{2} + \frac{1}{N}\sum_{\alpha=1}^{N} \sum_{\beta < \alpha} e^{-2i k (x_\alpha - x_\beta)}  \right| 
	\cdot \rho(\bm{\tau})
	\label{eq085}
\end{equation}
This value can be either larger or smaller than the exact spectral radius, therefore it can not be taken as an upper bound. However, usually it approximates more accurately the real spectral radius than the aforementioned upper bound based on  matrix--norms, see Eq.~\eqref{eq045}, which in general results more conservative. \\

The analysis carried out above for rods will be extended now to the more general case studied in this article: scatterers are generic resonators and longitudinal and flexural waves are coupled. The scattering coefficients are within matrices $\hat{\mathbf{G}}$ and $\hat{\mathbf{T}}$, defined in Eq. \eqref{eq046}. The block-matrices of $\hat{\mathbf{G}}$ are evaluations of the Green-matrix $\mathbf{G}(x)$ for every pair of scatterers. As developed in \ref{ap_green}, this matrix is a combination of the different modes traveling along the beam. Thus, for $x > 0$ we have
\begin{equation}
	\mathbf{G}(x) =  
	\mathbf{u}_1  \, \mathbf{v}_1^T  \, e^{q_1 x} +
	\mathbf{u}_2  \, \mathbf{v}_2^T  \, e^{q_2 x} +
    \mathbf{u}_3  \, \mathbf{v}_3^T  \, e^{q_3 x} 	\ , \qquad x \geq 0
    \label{eq109}
\end{equation}
where $q_1 = - i k_l$, $q_2 = - i k_b$ and $q_3 = - i k_s$ are three of the set of eigenvalues of matrix $\mathbf{A}$ and $k_l, \ k_b,$ and $k_s$ are wavenumbers of longitudinal, bending and shear waves, respectively (the latter for $\omega > \omega_c$). In an analogous way, we will first define (non-dimensional) matrix with relative impedance between the medium and the $\alpha$th scatterer, say
\begin{equation}
	\bm{\tau}_\alpha = \mathbf{G}(0^+) \, \mathbf{t}_\alpha = \mathbf{G}(0^+) \, \mathbf{K}_\alpha \, \left[ \mathbf{I} - \mathbf{G}(0^+) \mathbf{K}_\alpha\right] ^{-1}
	\label{eq107}
\end{equation}
Note the equivalent form, now under a matrix structure, between Eqs. \eqref{eq097} and \eqref{eq107}. According to Eq. \eqref{eq109}, the matrix $ \mathbf{G}(0^+)$ is defined as
\begin{equation}
	\mathbf{G}(0^+) =  
	\mathbf{u}_1  \, \mathbf{v}_1^T  +
	\mathbf{u}_2  \, \mathbf{v}_2^T   +
	\mathbf{u}_3  \, \mathbf{v}_3^T  
	\label{eq110}
\end{equation}
In \ref{ap_modes} an analytical form of vectors $\mathbf{u}_j$ and $\mathbf{v}_j$ can be found. The mechanism of propagation is more complex in this case, with coupling between the different wave modes. The medium presents interaction propagating waves with the distribution of scatterers.  Thus, we can generalize our proposed approximation of the spectral radius as
\begin{equation}
	\rho(\mathbf{g}) 
	\approx 
	\begin{cases}
		\displaystyle  \max  \left\{ 
	\rho \left[\bm{\mathcal{G}}(k_l) \right], \rho \left[\bm{\mathcal{G}}(k_b) \right] 
	\right\}
	\cdot \max_{1 \leq \alpha  \leq N} \rho(\bm{\tau}_\alpha) 	 & \text{if } \omega < \omega_c \\
	\displaystyle \max \left\{ 
	\rho  \left[\bm{\mathcal{G}}(k_l) \right], \rho \left[\bm{\mathcal{G}}(k_b) \right], \rho \left[\bm{\mathcal{G}}(k_s) \right] 
	\right\}
	\cdot \max_{1 \leq \alpha  \leq N} \rho(\bm{\tau}_\alpha) 	 & \text{if } \omega \geq \omega_c
	\end{cases}
	\label{eq106}
\end{equation}
where the matrix $\bm{\mathcal{G}}(k)$ has already been defined in Eq. \eqref{eq100} and $\rho(\bm{\tau}_\alpha) $ is the one introduced in Eq. \eqref{eq107}. The accuracy of this approximation depends (a) on the contrast of properties between scatterers and (b) on the correlation between the positions of scatterers. Thus, better results are obtained when all scatterers are equal and the distribution is random, with no special correlation patterns between positions which leads to noticeable periodicities. On the other side, periodic distribution and high contrast of parameters between resonators leads to worse results although in general the proposed approach overestimates the spectral radius. The above approximations for the spectral radius of the scattering matrix will be validated for two numerical examples covering longitudinal and flexural waves. \\

\begin{figure}[H]%
	\begin{center}
		\begin{tabular}{cc}
			\includegraphics[width=6.8cm]{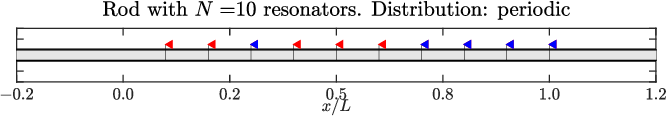} & \phantom{aa}
			\includegraphics[width=7.0cm]{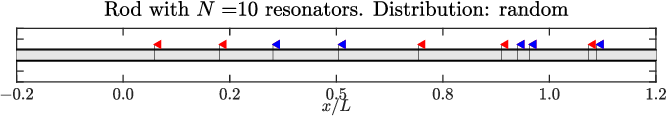} \\	\\		
			\includegraphics[width=7cm]{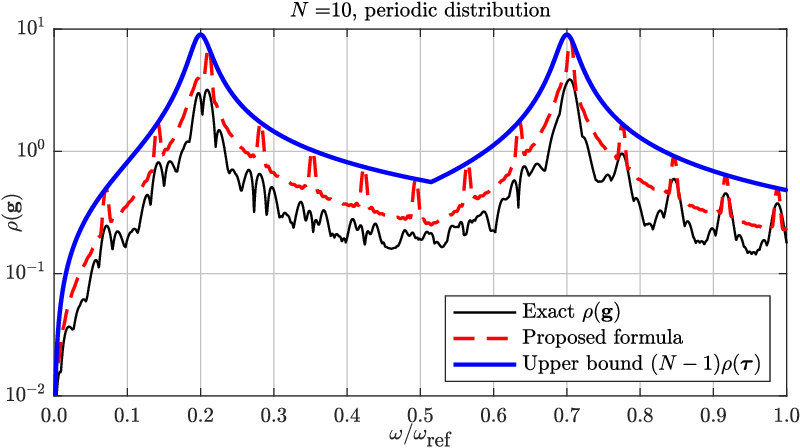} & 
			\includegraphics[width=7cm]{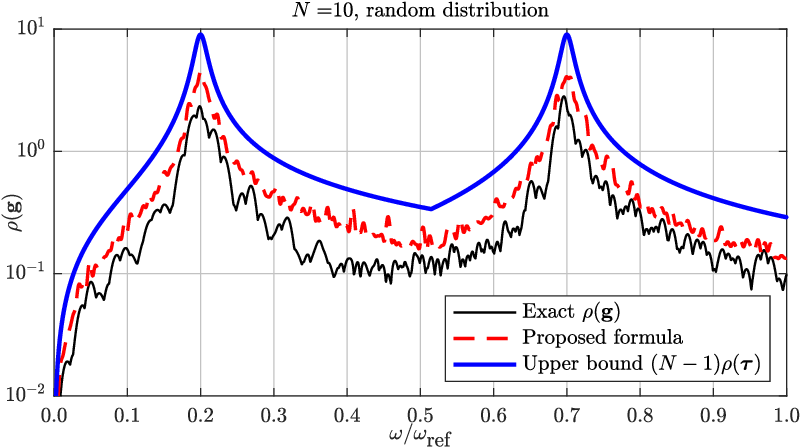} 			
		\end{tabular}
		\caption{A rod $EA =1.75 \times 10^5$ kN and mass per unit of length $\rho A = 5.25$ kg/m. $N=10$ resonators with mass $m_\alpha = 1.3 \times 10^{-2}$ kg and resonances $\omega_\alpha = 0.2	\omega_{\text{ref}} $ (red resonators) and $\omega_\alpha = 0.7	\omega_{\text{ref}} $ (blue resonators).  (left) periodic distribution of $N=10$ resonators; (right) random distribution of $N=10$ resonators. Black line: exact solution $\rho(\mathbf{g})$; Red-dashed line: approximate solution from Eqs.~\eqref{eq085} and Blue-continuous line: upper bound, $\rho_{\max{}}(\mathbf{g}) = (N-1) \rho(\bm{\tau})$}%
		\label{fig05}%
	\end{center}
\end{figure}

We consider first a rod with sectional stiffness $EA =1.75 \times 10^5$ kN and mass per unit of length $\rho A = 5.25$ kg/m. In Fig. \ref{fig05} we evaluate the spectral radius of the scattering matrix corresponding to $N = 10$ scatters with mass $m_\alpha = 1.3 \times 10^{-2}$ kg and resonances $\omega_\alpha = 28$ kHz (red scatters) and $\omega_\alpha = 8$ kHz (blue scatters). The range of validity will be studied in certain frequency range compatible with the classical rod model. Thus, in Ref. \cite{Graff-1999}, it is established a reasonable limit of validity of the classical rod model the frequency
\begin{equation}
	\omega_{\text{ref}} = 0.30 \sqrt{\frac{EA}{\rho I_x \nu^2}} \approx 40 \ \textnormal{kHz}
\end{equation}
where $I_x$ is the polar moment of inertia of the cross section and $\nu$ is the Poisson coefficient. The above limit is obtained comparing the classical rod model with the higher-order Love rod-model, establishing that in the range $0 \leq \omega \leq \omega_{\text{ref}}$ there exist good agreement between both. Hence,  we shall then consider a frequency range for the computations in this example as $ 0 \leq \omega / \omega_{\text{ref}} \leq 1$. In Fig. \ref{fig05} the spectral radius of the scattering matrix $\rho (\mathbf{g})$ for a periodic (Fig. \ref{fig05}--left) and a random distribution (Fig. \ref{fig05}--right) of scatterers has been plotted (continuous-black). The upper bound, computed as $  \rho_{\max{}} (\mathbf{g}) = (N-1)  \rho(\bm{\tau}) $ is plotted in continuous-blue line. Always between both curves emerge (dashed red line) the proposed approximate expression, given by Eq.~\eqref{eq085}. In both cases the intensity of the scattering around the two resonant frequencies of the system is perceived. It is observed that the proposed model fits this effect always overestimating the real value, which allows us to predict the range of behavior in weak scattering safely. In the periodic case, the separation between scatters is $s = 1$ m, thus the first Bragg scattering peak emerges approximately at $\omega_{\text{Bragg}} = \pi c/s \approx 0.07 \omega_{\text{ref}}$. The proposed formula proposed   satisfactorily predicts such set of peaks due to periodicity in Fig. \ref{fig05} (left), the effect of which is accentuated at high frequencies. In Fig. \ref{fig05} (right) the case of $N=10$ randomly distributed scatterers is shown. Still in this case, the prediction lies above the real value but not trespassing the upper bound. \\

\begin{figure}[H]%
	\begin{center}
		\begin{tabular}{ccc}
			\multicolumn{3}{c}{\includegraphics[width=8cm]{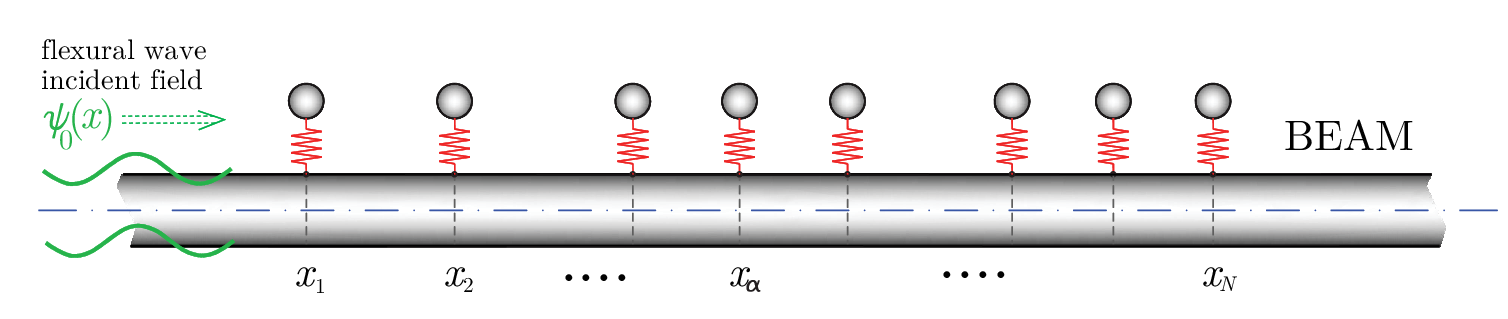}  } \\	\\		
			\includegraphics[width=7cm]{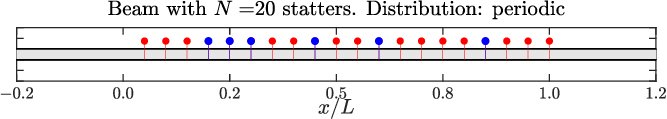} & \phantom{aaa} &
			\includegraphics[width=7cm]{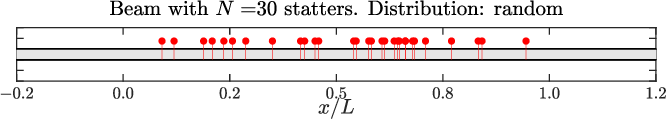} \\	\\		
			\includegraphics[width=7cm]{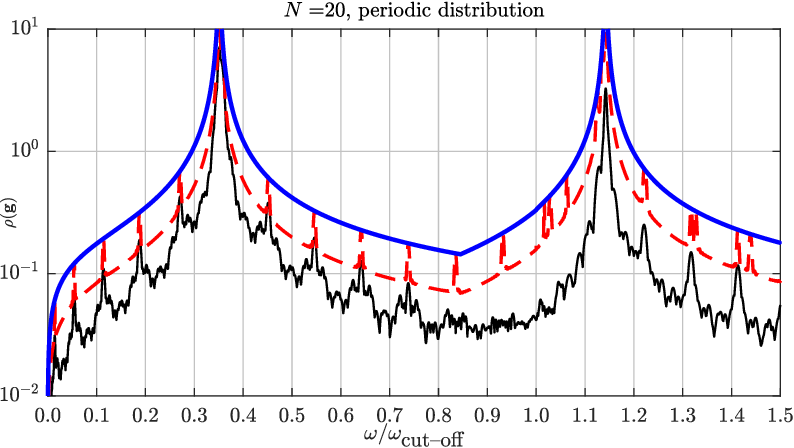} & \phantom{a} &
			\includegraphics[width=7cm]{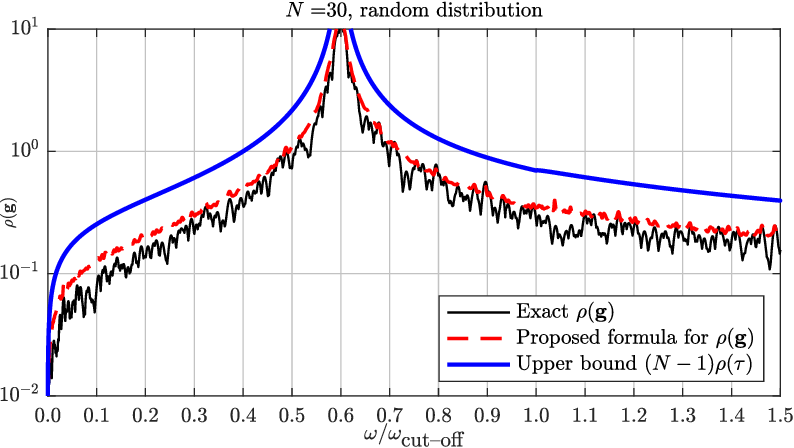} 			
		\end{tabular}
		\caption{Spectral radius distribution with frequency for a beam. (Left) periodic distribution of $N=20$ resonators; 6 of them with resonance of $\omega_\alpha = 0.35 \omega_c$ and the rest with  $\omega_\alpha = 1.14 \omega_c$. (Right) random distribution of $N=30$ equal resonators with resonance $\omega_\alpha = 0.60 \omega_c$. Black line: exact solution $\rho(\mathbf{g})$; Red-dashed line: approximate solution, Eqs.~\eqref{eq084} and \eqref{eq106} ; Blue-continuous line: upper bound, $\rho(\mathbf{g}) \leq (N-1) \rho(\bm{\tau})$}%
		\label{fig06}%
	\end{center}
\end{figure}

Secondly, we consider a Timoshenko beam with stiffness parameters $EI = 36$ m$^2$kN, $GA_z = 5738$ kN and inertia parameters per unit of length $\rho A = 0.525$ kg/m and $\rho I_y = 1.1 \times 10^{-6}$ m$^2$kg/m. The cut-off frequency is $\omega_c = 364$ kHz.  
In Fig. \ref{fig06}(left) a periodic distribution of $N=20$ point-resonators with mass $m_\alpha = 0.1$ kg have been considered. Among this set, 14 of them have a resonance of 
$\omega_\alpha / \omega_c = 0.35$, and the rest (chosen randomply) $\omega_\alpha / \omega_c = 1.14$. It can be observed clearly the almost equidistant peaks corresponding to the Bragg resonances: (i) below the cut-off frequency the peaks location are the solution of the equations $k_b(\omega) = \pi / s, \ 2 \pi / s, 3 \pi / s, \ldots$, where $k_b(\omega)$ is the dispersion relationship of bending waves and $s$ is the separation between scatterers. (ii) above the cut-off frequency we can distinguish two families of Bragg peaks, that one corresponding to the bending waves and a new one corresponding to shear waves, solution of equations $k_s(\omega) = \pi / s, \ 2 \pi / s, 3 \pi / s, \ldots$, where $k_s(\omega)$ is the dispersion relationship of shear waves. Bragg peaks correspond to maximum values of eigenvalues of matrices $\left[\bm{\mathcal{G}}(k_b) \right] $ and $\rho \left[\bm{\mathcal{G}}(k_b) \right] $, and they are independent of the type of scatterers. On the other hand, the resonances are singularities of the matrix $\rho(\bm{\tau})$ and they emerge in the plot of spectral radius as vertical asymptotes, defining points around which clearly weak-scattering approximations cannot be assumed. \\

In Fig. \ref{fig05}(right) a random distribution of $N=30$ point-resonators have been considered, all with the same single resonance of $\omega_\alpha / \omega_c = 0.60$. Since there are not apparent correlation between positions of scatterers,  there are no evidences of Bragg peaks. However, it is clear the location of the resonance, since it does not depend on the positions. In this case, the approximate model estimates much more accurately the real value of the spectral radius. \\

\subsection{Closed form expressions for Born approximation}

\begin{figure}[h]%
	\begin{center}
		\includegraphics[width=15cm]{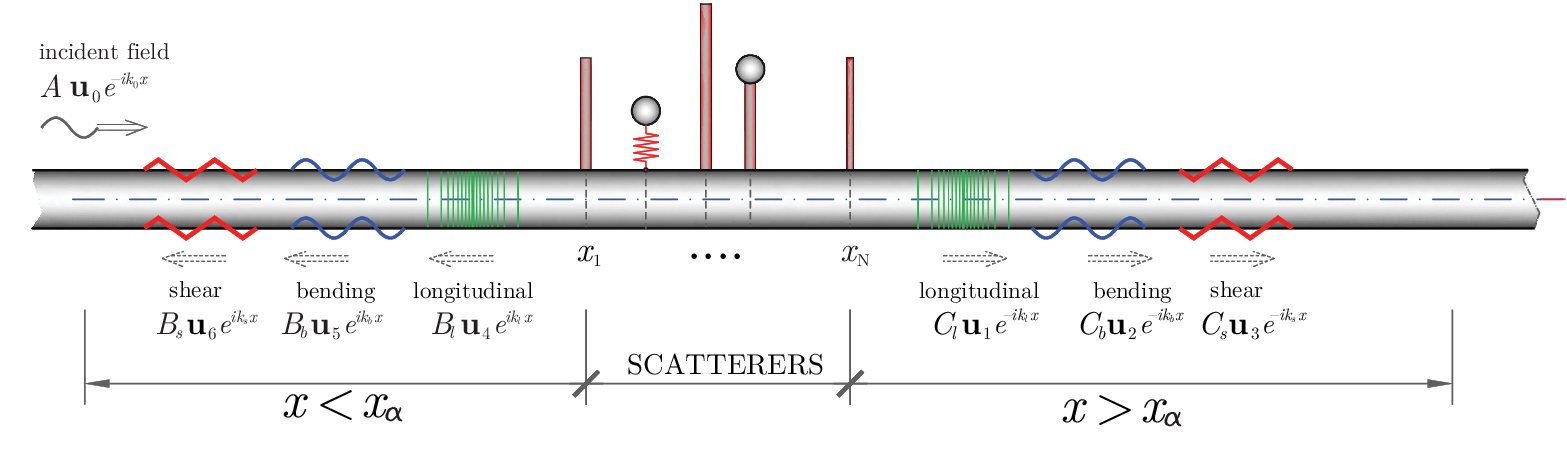} \\	
		\caption{Beam-rod with an arrangement of $N$ resonators. Reflection ($B_l, B_b, B_s$) and transmission ($C_l, C_b, C_s$) corresponding to each one of the three modes propagating, for an incident field given by $ \bm{\psi}_0(x) = A \, \mathbf{u}_0 \, e^{- i k_0 \, x}$  }%
		\label{fig12}%
	\end{center}
\end{figure}

Assuming that the scattering level is low enough to consider the weak-scattering approximations and the Neumann series, let us derive closed form expressions from the Born approximations at any point in the far-field from the cluster of scatterers. Consider any incident wave coming from the left and righwards. In general, it will have the form
\begin{equation}
	 \bm{\psi}_0(x) = A \, \mathbf{u}_0 \, e^{- i k_0 \, x}
	 \label{eq111}
\end{equation}
where $A$ is a coefficient and  $\{-i k_0, \mathbf{u}_0 \}$ represents a propagating eigenpair (eigenvalue and eigenvector) of matrix $\mathbf{A}$. The Born approximation of the scattered field is then
\begin{equation}
	\mathbf{u}_s(x) = \sum_{\alpha=1}^N \mathbf{G}(x-x_\alpha) \,  \mathbf{t}_\alpha \, \bm{\psi}_0(x_\alpha) = 
										A \, \sum_{\alpha=1}^N \mathbf{G}(x-x_\alpha) \,  \mathbf{t}_\alpha \, \mathbf{u}_0 \, e^{- i k_0 \, x}  \label{eq112}
\end{equation}
Let us assume that the frequency is $\omega \geq \omega_c$, so we will obtain certain expression involving all modes. In the low-frequency case, the terms corresponding to shear waves can be neglected at the far-field, since these modes are evanescent. \\

\begin{table}
	\begin{center}
		\begin{tabular}{|c|ccc|ccc|}
	\hline
	Direction of propagation	&	\multicolumn{3}{c|}{Rightwards, $\rightarrow$}  & \multicolumn{3}{c|}{Leftwards, $\leftarrow$} \\
	\hline
	Mode	&	 1 (long.)&	2 (bending)	&3 (shear)&	4 (long.)&	5 (bending)	&	6 (shear.)			\\
	Eigenvector &		$\mathbf{u}_1$ & 		$\mathbf{u}_2$ &		$\mathbf{u}_3$ &		$\mathbf{u}_4$ &		$\mathbf{u}_5$ &		$\mathbf{u}_6$  \\
		\hline
	Eigenvalue & $q_1$ 			&			$q_2$ 	&			$q_3$ 	&			$q_4$ 			&			$q_5$ 			&			$q_6$					\\
	$\omega < \omega_c$		&			$-i k_l$		&  			$-i k_b$		 & 			$- k_s$		 	&			$+i k_l$		 & 			$+ik_b$		 & 			$+k_s$		 \\
	$\omega > \omega_c$		&			$-i k_l$		&  			$-i k_b$		 & 			$-i k_s$		 & 			$+i k_l$		 & 			$+ik_b$		 & 			$+ ik_s$		 \\
	\hline
\end{tabular}
	\end{center}
	\caption{Modes propagation of 1D waveguides}
	\label{tab01}
\end{table}

Consider first any point to the left of the cluster of scatterers, i.e. $x < x_\alpha$, for $1 \leq \alpha \leq N$. Using the closed form expression of the Green function derived in \ref{ap_green}, with the main results in Table \ref{tab01}, and reminding that we are considering $\omega \geq \omega_c$, it yields
\begin{eqnarray}
	\mathbf{G}(x - x_\alpha) 
	&=& -\sum_{j = 4}^{6}  \, \mathbf{u}_j  \, \mathbf{v}_j^T   \, e^{q_j x}  \nonumber \\
	&=& -     \mathbf{u}_4\, \mathbf{v}_4^T e^{i k_l \, (x - x_\alpha)}  
	-    \mathbf{u}_5 \, \mathbf{v}_5^T  \,  e^{i \, k_b (x - x_\alpha)} 
	-    \mathbf{u}_6 \, \mathbf{v}_6^T \, e^{i k_s (x - x_\alpha)}  \ , \quad x < x_\alpha
	\label{eq113}
\end{eqnarray}
Plugging this expression into Eq.~\eqref{eq112} leads to
\begin{eqnarray}
	\mathbf{u}_s(x) &=& \sum_{\alpha=1}^N \mathbf{G}(x-x_\alpha) \,  \mathbf{t}_\alpha \, \bm{\psi}_0(x_\alpha) = 
	A \, \sum_{\alpha=1}^N \mathbf{G}(x-x_\alpha) \,  \mathbf{t}_\alpha \, \mathbf{u}_0 \, e^{- i k_0 \, x_\alpha}  \nonumber \\ 
		&=& A \, \sum_{\alpha=1}^N 
		\left[  -     \mathbf{u}_4\, \mathbf{v}_4^T e^{i k_l \, (x - x_\alpha)}  
		-    \mathbf{u}_5 \, \mathbf{v}_5^T  \,  e^{i \, k_b (x - x_\alpha)} 
		-    \mathbf{u}_6 \, \mathbf{v}_6^T \, e^{i k_s (x - x_\alpha)}  
		\right] \,  \mathbf{t}_\alpha \, \mathbf{u}_0 \, e^{- i k_0 \, x_\alpha} \nonumber \\
		&\equiv & B_l \, \mathbf{u}_4  e^{i k_l \, x}   + B_b \, \mathbf{u}_5  e^{i k_b \, x}    + B_s \, \mathbf{u}_6  e^{i k_s \, x}   
		\ , \quad x < x_1 <  \cdots < x_N
		 \label{eq114}
\end{eqnarray}
We observe that the reflected wave (scattered wave leftwards) propagates under the contribution of the three main modes of the guide when the incident wave mode is $\mathbf{u}_0  e^{- i k_0 \, x}$. The third term in Eq.~\eqref{eq114} can be neglected for low-frequency case ($\omega < \omega_c$). 
The coefficients associated to each mode, $B_l$ (longitudinal wave), $B_b$ (bending wave) and $B_s$ (shear wave) respect to the incident coefficient $A$ are
\begin{eqnarray}
	B_l/A &=& - \, \sum_{\alpha = 1}^N \, \mathbf{v}_4^T \, \mathbf{t}_\alpha  \, \mathbf{u}_0 \, e^{-i (k_0 + k_l) \, x_\alpha} \nonumber \\
	B_b/A &=& - \, \sum_{\alpha = 1}^N \, \mathbf{v}_5^T \, \mathbf{t}_\alpha  \, \mathbf{u}_0 \, e^{-i (k_0 + k_b) \, x_\alpha} \nonumber \\
	B_s/A &=& - \, \sum_{\alpha = 1}^N \, \mathbf{v}_6^T \, \mathbf{t}_\alpha  \, \mathbf{u}_0 \, e^{-i (k_0 + k_s) \, x_\alpha} 
	 \label{eq115}	
\end{eqnarray}
If matrix $\mathbf{T}_\alpha  $ contains coupling terms between the propagating modes, then the above coefficients will be distinct of zero and any incident wave will be radiated back in a wave which includes every admissible wavemode in the medium. \\

Let us consider secondly a point located to the right of the scatterers and assuming as before that $\omega > \omega_c$ in order to visualize all modes. Then the Green function for  $x > x_\alpha$, for $1 \leq \alpha \leq N$ can be written as
\begin{eqnarray}
	\mathbf{G}(x - x_\alpha) 
	&=& \sum_{j = 1}^{3}  \, \mathbf{u}_j  \, \mathbf{v}_j^T   \, e^{q_j x}  \nonumber \\
	&=&      \mathbf{u}_1\, \mathbf{v}_1^T e^{ -i k_l \, (x - x_\alpha)}  
			    + \mathbf{u}_2 \, \mathbf{v}_2^T  \,  e^{-i \, k_b (x - x_\alpha)} 
	+    \mathbf{u}_3 \, \mathbf{v}_3^T \, e^{-i k_s (x - x_\alpha)}  \ , \quad x > x_\alpha
	\label{eq116}
\end{eqnarray}
Evaluating Eq.~\eqref{eq112} for the above expression, it yields
\begin{eqnarray}
	\mathbf{u}_s(x) &=& \sum_{\alpha=1}^N \mathbf{G}(x-x_\alpha) \,  \mathbf{t}_\alpha \, \bm{\psi}_0(x_\alpha) = 
	A \, \sum_{\alpha=1}^N \mathbf{G}(x-x_\alpha) \,  \mathbf{t}_\alpha \, \mathbf{u}_0 \, e^{- i k_0 \, x_\alpha}  \nonumber \\ 
	&=& A \, \sum_{\alpha=1}^N 
	\left[  \mathbf{u}_1\, \mathbf{v}_1^T e^{ -i k_l \, (x - x_\alpha)}  
	+ \mathbf{u}_2 \, \mathbf{v}_2^T  \,  e^{-i \, k_b (x - x_\alpha)} 
	+    \mathbf{u}_3 \, \mathbf{v}_3^T \, e^{-i k_s (x - x_\alpha)}  
	\right] \,  \mathbf{t}_\alpha \, \mathbf{u}_0 \, e^{- i k_0 \, x_\alpha} \nonumber \\
	&\equiv & C_l \, \mathbf{u}_1  e^{- i k_l \, x}   + C_b \, \mathbf{u}_2  e^{-i k_b \, x}    + C_s \, \mathbf{u}_3  e^{-i k_s \, x}   
	\ , \quad x > x_N >  \cdots > x_1
	\label{eq117}
\end{eqnarray}
The coefficients of the associated to the scattered wavemodes for $x > x_N$ are
\begin{eqnarray}
	C_l/A &=&  \, \sum_{\alpha = 1}^N \, \mathbf{v}_1^T \, \mathbf{t}_\alpha  \, \mathbf{u}_0 \, e^{-i (k_0 - k_l) \, x_\alpha} \nonumber \\
	C_b/A &=& \, \sum_{\alpha = 1}^N \, \mathbf{v}_2^T \, \mathbf{t}_\alpha  \, \mathbf{u}_0 \, e^{-i (k_0 - k_b) \, x_\alpha} \nonumber \\
	C_s/A &=&  \, \sum_{\alpha = 1}^N \, \mathbf{v}_3^T \, \mathbf{t}_\alpha  \, \mathbf{u}_0 \, e^{-i (k_0 - k_s) \, x_\alpha} 
	\label{eq115b}	
\end{eqnarray}
The wavefield so calculated corresponds to the first-order solution in terms of the scattering coefficients, i.e. $\mathbf{u}_s(x)$ is linear in the coefficients of $\mathbf{K}_\alpha$. The exponential terms vanish when the mode of the transmission coefficient coincides with that of the incident mode, because in that case $k_0 = k_\text{scattered}$. The lost of accuracy becomes important since there is no information about the relative positions between scattered. Higher order terms need to be taken in order to capture the relative positions influence in the coefficients $C_l, \ C_b$ and $C_s$. This makes the Born-approximation of the transmitted wavefield always less accurate than the reflected wave, as will be shown in the numerical examples. 

\section{Numerical examples}

\begin{figure}[H]%
	\begin{center}
		\begin{tabular}{cc}
			\includegraphics[width=7.3cm]{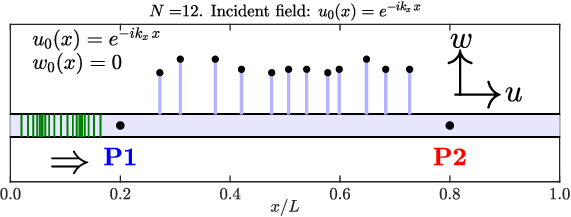} & 			
			\includegraphics[width=8.0cm]{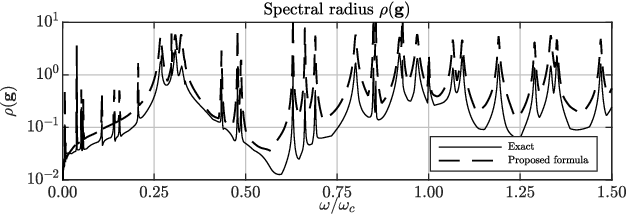} \\ 		
			\includegraphics[width=8.0cm]{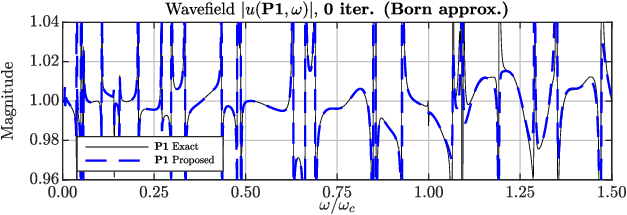} & 
			\includegraphics[width=8.0cm]{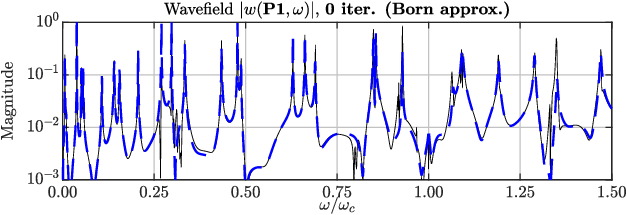} \\ 
			\includegraphics[width=8.0cm]{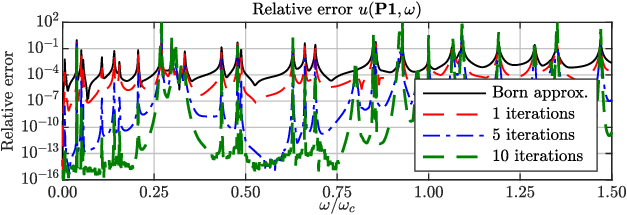} & 
			\includegraphics[width=8.0cm]{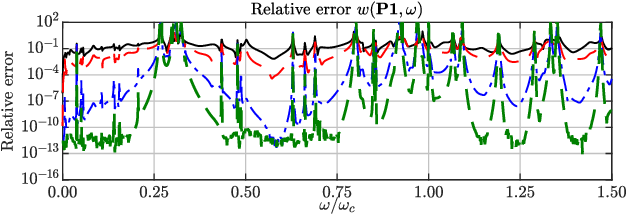} \\  
			\includegraphics[width=8.0cm]{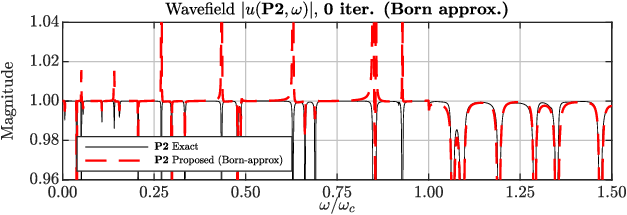} &
			\includegraphics[width=8.0cm]{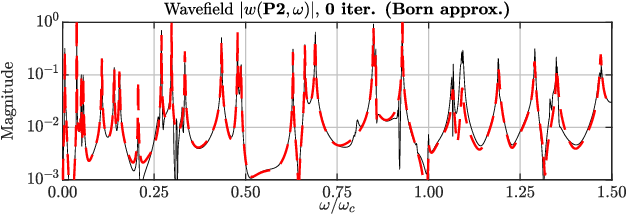} \\  			\\
			\includegraphics[width=8.0cm]{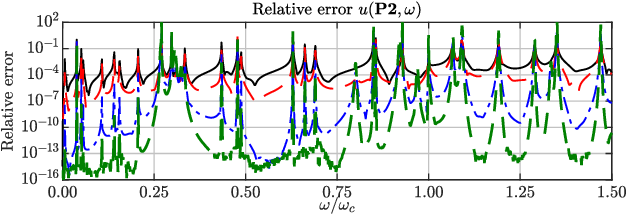} &
			\includegraphics[width=8.0cm]{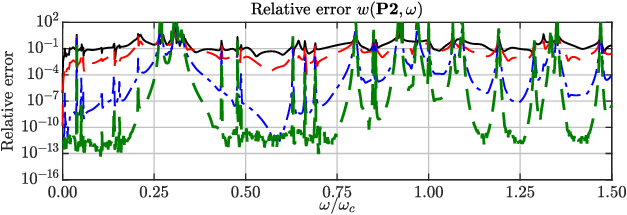} \\ 					
		\end{tabular}
		\caption{}%
		\label{fig07}%
	\end{center}
\end{figure}

In this section some numerical examples will help to validate and interpret the results obtained in this article. Our main objective is to study the range of validity of the Born approximation for coupled longitudinal and bending waves in a waveguide with multiple resonators in the form of small attached beams. Fig. \ref{fig07} shows the results for an aluminum beam of cross section 12$\times$12 cm$^2$ and a set of $N=12$ resonators of section 1.5$\times$1.5 cm$^2$, randomly distributed. Three different lengths have been considered for the resonators: $l_\alpha = \{34, 36, 41\}$ cm, which give rise to a varied spectrum of vibration frequencies both in the longitudinal and in the transverse direction of the resonator. These frequencies can be observed both in the peaks of the reflected wavefield and in the spectral radius representation. In Fig. \ref{fig07} (top-right) the spectral radius of the scattering matrix $\rho (\mathbf{g})$ has been plotted: the main indicator of the applicability of the Born approximation. In addition, we have also plotted an estimation of this parameter using the approximate formula proposed in Eq. \eqref{eq106}, which, as predicted by the theoretical results, approaches the exact value even at the peaks corresponding to the resonance frequency, but in general overestimating it. In the sketch of the structure, Fig. \ref{fig07} (top-left), the longitudinal incident wave has been depicted. The set of scatterers will produce a reflection of the wave, $u(x,\omega)$, and in addition bending waves will be radiated by the excitation of the transverse modes of vibration of the resonators, giving rise to the field $w(x,\omega)$. The wavefield magnitudes $\left|u(x,\omega) \right|$ and $\left|w(x,\omega) \right|$ evaluated with the Born approximation have been plotted in the frequency domain at points $\textbf{P1}$ (before the scatterers) and $\textbf{P2}$ (after).

In general, the reflected wave (point \textbf{P1}) by the set of scatterers can be evaluated with the Born approximation over a wide range of frequencies. However, the transmitted wave (point \textbf{P2}) has larger errors and is more sensitive to scatter resonances. The evaluation of more iterations in the iterative scheme of Eq. \eqref{eq088} leads to results in accordance with those predicted by the theory: in regions with $\rho (\mathbf{g})<1$ it converges and the error drops rapidly after a few iterations, while in those frequencies with $\rho (\mathbf{g})\geq 1$, divergence is manifested.\\

\begin{figure}[H]%
	\begin{center}
		\begin{tabular}{cc}
			\includegraphics[width=7.3cm]{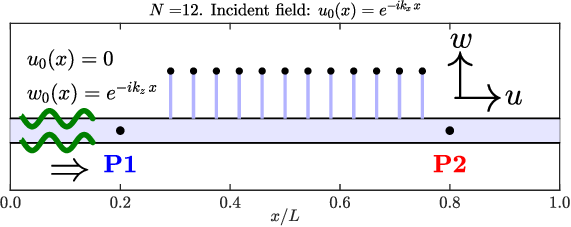} & 			
			\includegraphics[width=8.0cm]{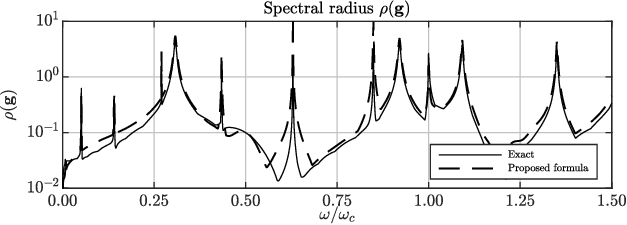} \\ 		
			\includegraphics[width=8.0cm]{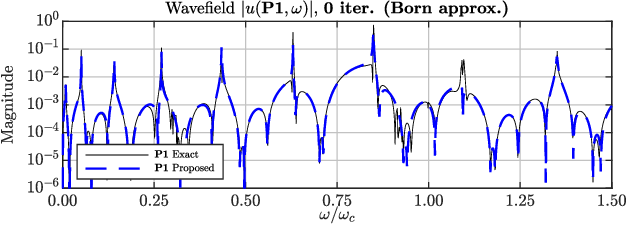} & 
			\includegraphics[width=8.0cm]{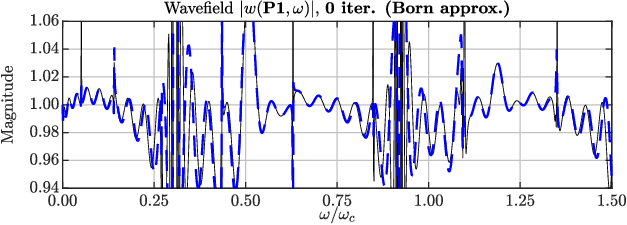} \\ 
			\includegraphics[width=8.0cm]{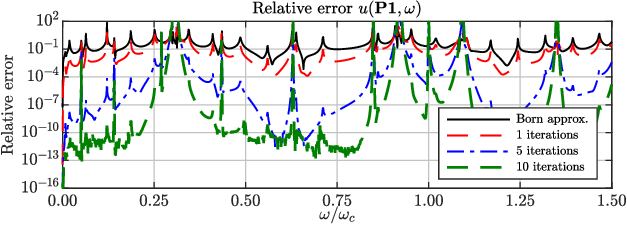} & 
			\includegraphics[width=8.0cm]{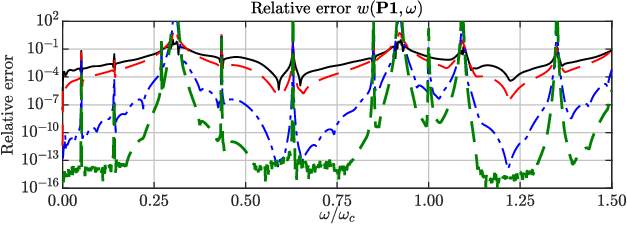} \\  
			\includegraphics[width=8.0cm]{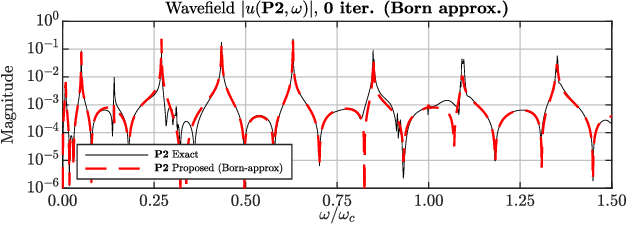} &
			\includegraphics[width=8.0cm]{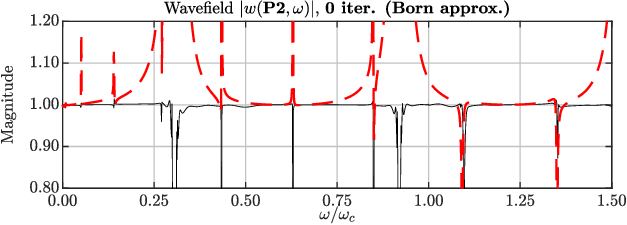} \\  			
			\includegraphics[width=8.0cm]{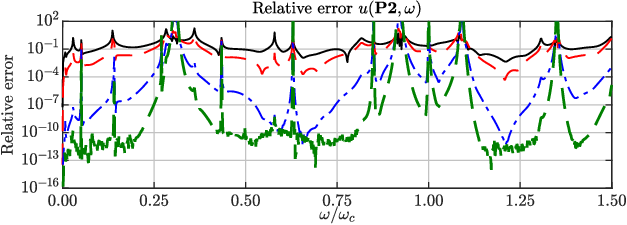} &
			\includegraphics[width=8.0cm]{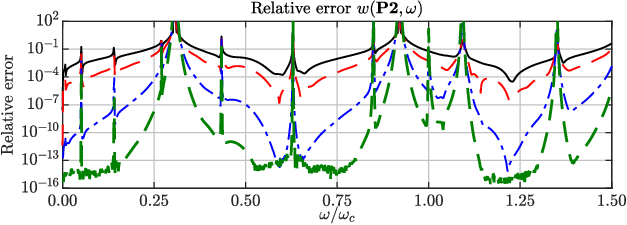} \\ 					
		\end{tabular}
		\caption{}%
		\label{fig08}%
	\end{center}
\end{figure}

Fig. \ref{fig08} shows the results of the simulations for the same host-beam but with a set of $N=12$ periodically distributed equal resonators of length $l_\alpha=36$ cm. The Bragg resonances associated with the periodicity are added to the set of scatterer resonances. The latter produce weaker scattering and their influence extends to a narrow band, generating sharp peaks. A greater presence of resonator bending vibration frequencies is manifested in the longitudinal wave response. On the other hand, although less numerous, the longitudinal modes of the oscillators affect the incident wave in a wider frequency range around such resonances. The Born approximation in general accurately predicts the reflected waves but not the wavefield behind the scatterer array, thus the transmitted response is poorer. More iterations are then needed to achieve the same accuracy.

\begin{figure}[H]%
	\begin{center}
		\begin{tabular}{crr}
			\includegraphics[width=8cm]{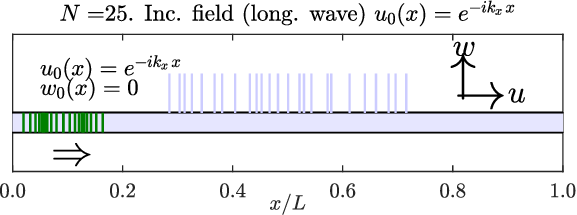}  &
			\includegraphics[width=8cm]{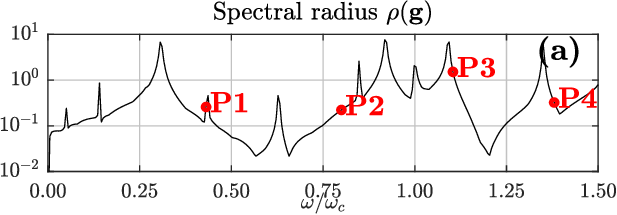} \\ \\
			\includegraphics[width=8cm]{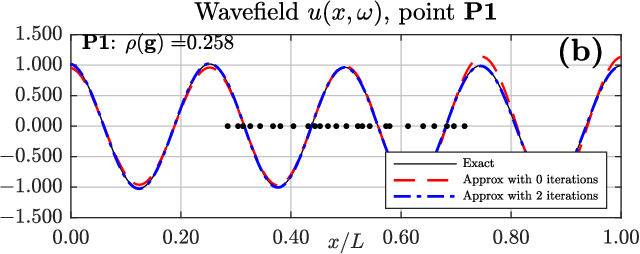} &
			\includegraphics[width=8cm]{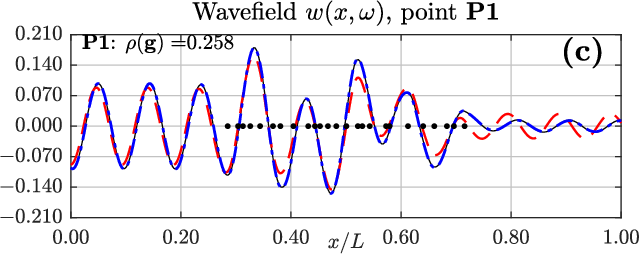}  \\  \\
			\includegraphics[width=8cm]{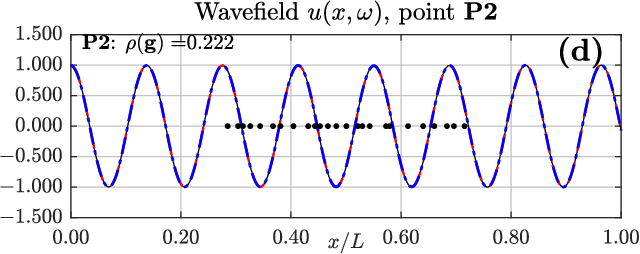} &
			\includegraphics[width=8cm]{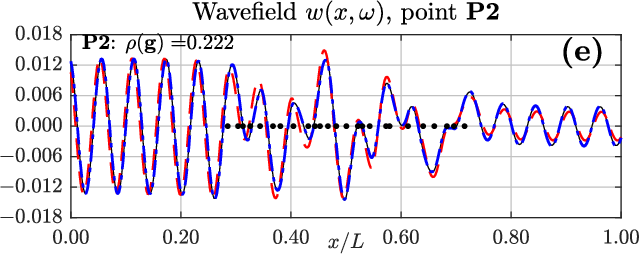}  \\  \\
			\includegraphics[width=8cm]{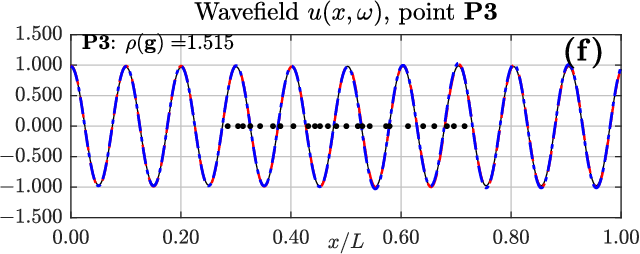} &
			\includegraphics[width=8cm]{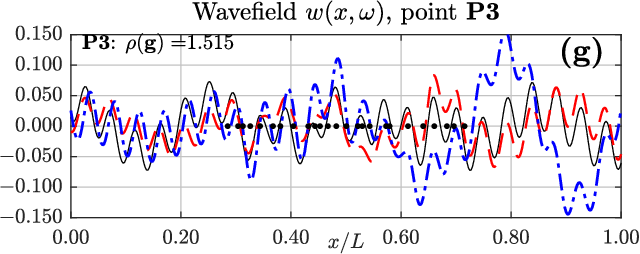}  \\  \\
			\includegraphics[width=8cm]{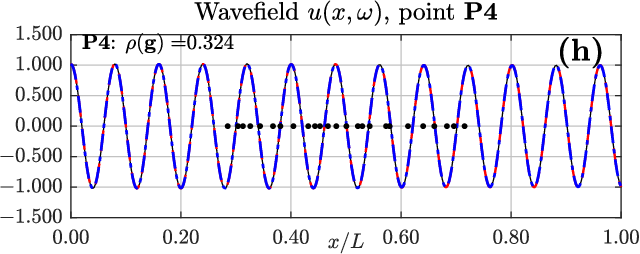} &
			\includegraphics[width=8cm]{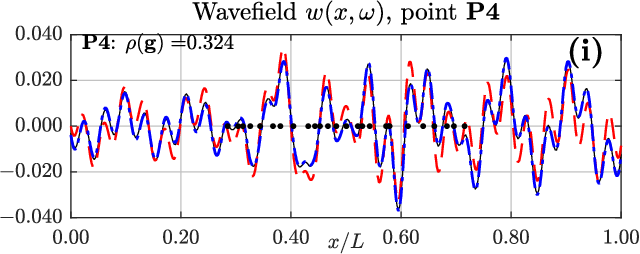}  \\ 
		\end{tabular}
		\caption{}%
		\label{fig09}%
	\end{center}
\end{figure}

\begin{figure}[H]%
	\begin{center}
		\begin{tabular}{crr}
			\includegraphics[width=8cm]{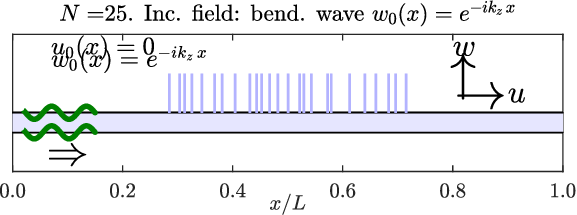}  &
			\includegraphics[width=8cm]{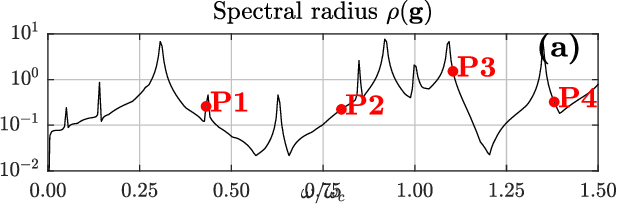} \\ \\
			\includegraphics[width=8cm]{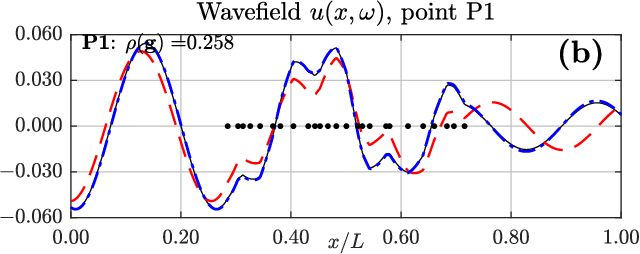} &
			\includegraphics[width=8cm]{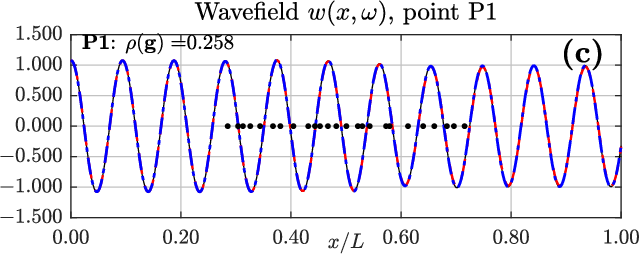}  \\ \\
			\includegraphics[width=8cm]{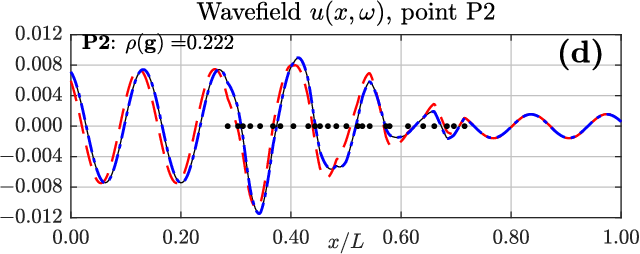} &
			\includegraphics[width=8cm]{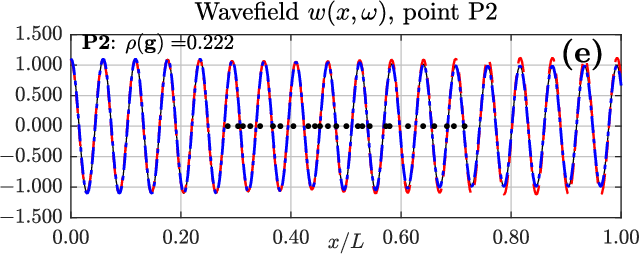}  \\ \\
			\includegraphics[width=8cm]{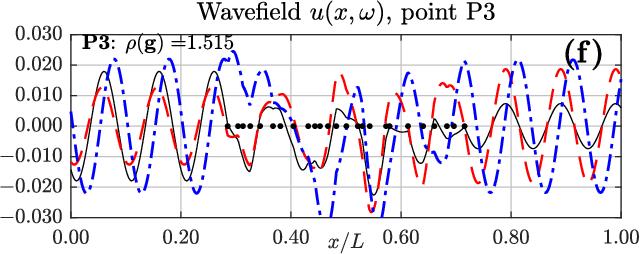} &
			\includegraphics[width=8cm]{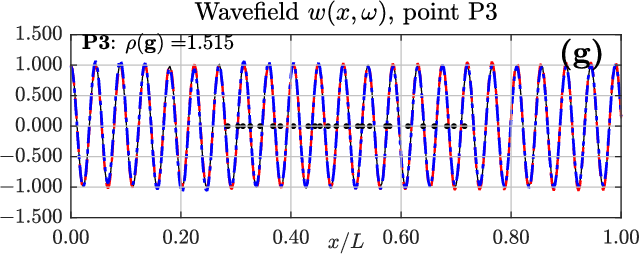}  \\ \\
			\includegraphics[width=8cm]{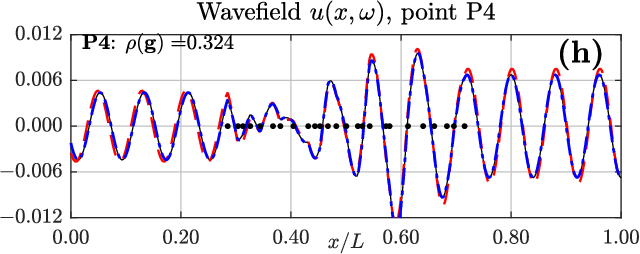} &
			\includegraphics[width=8cm]{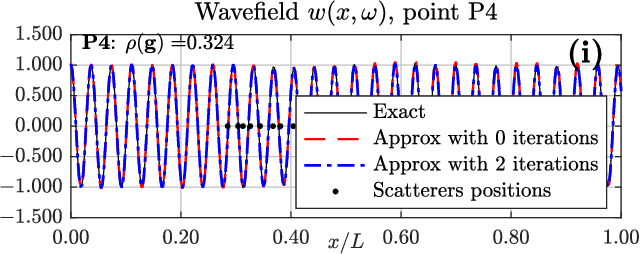}  \\ 
		\end{tabular}
		\caption{}%
		\label{fig10}%
	\end{center}
\end{figure}

In Figs. \eqref{fig09} and \eqref{fig10} the wavefield over the entire distance considered in a random configuration of $N=25$ resonators has been plotted. Fig. \eqref{fig09} shows the results for a longitudinal incident wave and Fig. \eqref{fig10} for a bending wave. The waveforms are represented for several frequencies distributed over the range considered, plotting in total 4 points (\textbf{P1} to \textbf{P4}). The simulations performed on the frequencies associated with weak scattering fit the exact field with high accuracy even with the Born approximation. The wave radiated by the cluster of scatterers manifests a more irregular shape as it propagates in both directions, with an amplitude a few orders of magnitude smaller than the incident wave. 

\begin{figure}[H]%
	\begin{center}
		\begin{tabular}{crr}
			&
			\includegraphics[width=5.3cm]{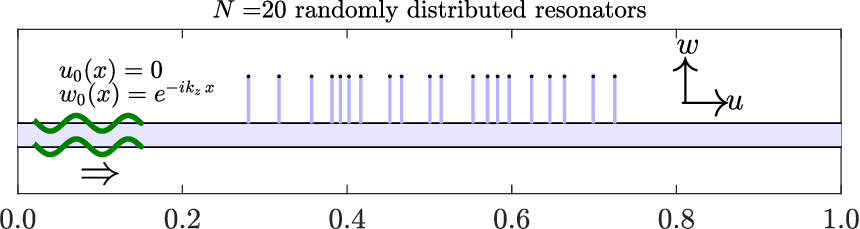} & 
			\includegraphics[width=5.3cm]{figures/figure_space-freq-map_BeamSketch.eps} \\ 
			\includegraphics[width=3.6cm]{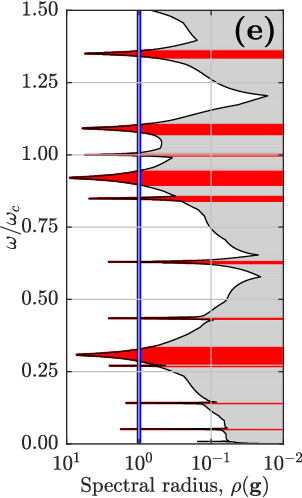} & 
			\includegraphics[width=6.1cm]{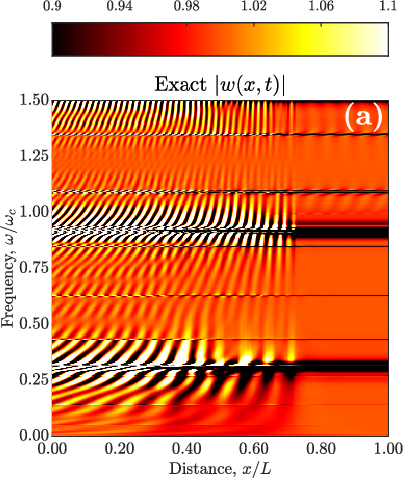} & 					
			\includegraphics[width=6.1cm]{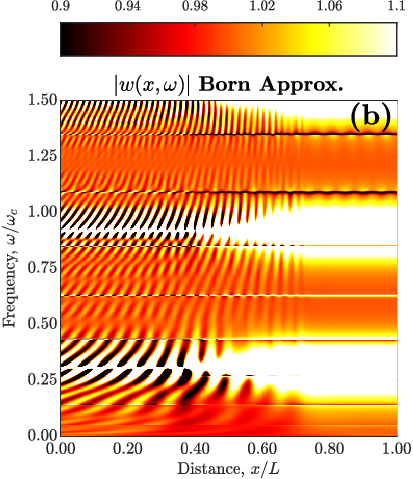} \\ 
			\includegraphics[width=3.6cm]{figures/figure_space-freq-map_SpectralRadius.eps} &								
			\includegraphics[width=6.1cm]{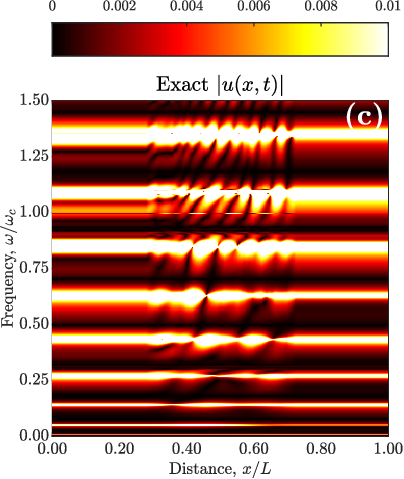} & 
			\includegraphics[width=6.1cm]{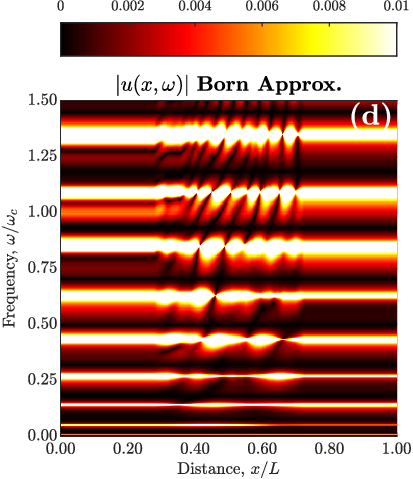} \\ 
		\end{tabular}
		\caption{}%
		\label{fig11}%
	\end{center}
\end{figure}

In Fig. \ref{fig11} the wave field obtained with the Born approximation for the whole wave range has been plotted on a map representing the magnitude of the transversal wave $\left|w(x,\omega)\right|$ and the longitudinal wave $\left|u(x,\omega)\right|$ in response to an incident bending wave $w_0(x) = e^{-ik_z x}$. A number of  $N=20$ randomly distributed scatterers are considered. Together with the numerical results, the spectral radius has been plotted Figs. \ref{fig11}(e)  as well as the regions where its value exceeds the limit of unity. In Fig. \ref{fig11}(b) the loss of accuracy of the transmitted wavefield after passing through the resonators cluster is clearly seen. However, the radiated longitudinal wave can be satisfactorily approximated over the entire frequency range and along the whole length of the system, as seen in Figs. \ref{fig11}(c) and \ref{fig11}(d). 

The simulated numerical examples have allowed to validate the estimation of the wavefield given by the Born approximation. The spectral radius has proved to be a valuable indicator of the quality of the approximation and of the convergence of the iterative procedure. However, the Born approximation of the wavefield transmitted through the scatterers is not accurately estimated. As reflected in Eqs. \eqref{eq115b}, the Born approximation of a mode loses the phase shift information when passing through an obstacle. Thus, for example, if the incident wave is longitudinal $k_0 \equiv k_l$ and $\mathbf{u}_0 \equiv \mathbf{u}_1$, the transmission coefficients of the three modes are.
\begin{eqnarray}
	C_l/A &=&  \, \sum_{\alpha = 1}^N \, \mathbf{v}_1^T \, \mathbf{t}_\alpha  \, \mathbf{u}_1 \nonumber \\
	C_b/A &=& \, \sum_{\alpha = 1}^N \, \mathbf{v}_2^T \, \mathbf{t}_\alpha  \, \mathbf{u}_1 \, e^{-i (k_l - k_b) \, x_\alpha} \nonumber \\
	C_s/A &=&  \, \sum_{\alpha = 1}^N \, \mathbf{v}_3^T \, \mathbf{t}_\alpha  \, \mathbf{u}_1 \, e^{-i (k_l - k_s) \, x_\alpha} 
	\label{eq122}	
\end{eqnarray}
Therefore, the information about the positions of the scatterers in the evaluation of the transmitted longitudinal wave $C_l$ is lost, although the coefficients $C_b$ and $C_s$ do present a dependency of such positions. This fact would explain why the radiated longitudinal wave $u(x,\omega)$ in Figs. \ref{fig11}(c) and \ref{fig11}(d) and excited by a bending incident wave $w_0(x) = e^{- i k_b x}$, reproduces much better the exact result than the transmitted transversal wave $w(x,\omega)$, shown in Figs. \ref{fig11}(a) and \ref{fig11}(b). In this case, it would be required to consider more terms in the Neumann series in order to capture the effect of the phase shift.

\section{Conclusions}

In this paper, elastic waveguides with an arbitrary set of obstacles with resonances are considered. The main objective is to rigorously discuss under which conditions the oscillator array induces weak scattering. A detailed study of the general form of the wavefield  for any type of resonators together with the multiple scattering equations are derived. The conditions to guarantee weak scattering are closely linked to the study of the spectral radius of the so-called scattering matrix. Under these conditions the Born approximation for elastic waves is proposed, whose application allows the estimation of the wave field under weak scattering conditions. Several numerical examples allow to validate the developments carried out over the whole frequency range.

\appendix

\section{Modes of 1D waveguides}
\label{ap_modes}

In this paper we are considering 1D waveguides of longitudinal and flexural waves, governed by the equation
\begin{equation}
	\frac{\textrm{d} \mathbf{u}}{\textrm{d} x} = \mathbf{A} \, \mathbf{u}
	\label{eq_AP01a}
\end{equation}
where 	$\mathbf{u}(x) = \{u,  w,  \theta,  N,  V,  M\}^T$ denotes the state-vector and $\mathbf{A}$ is the matrix defined in Eq.~\eqref{eq004b}. The general solution of the equation is formed by a superposition of the propagating modes along the beam. Denoting by  $\{q_j, \mathbf{u}_j\},  \ 1 \leq j \leq 6\}$ to the 6 eigenvalues and eigenvectors of matrix $\mathbf{A}$ then the general form of the force--free propagating motion is
\begin{equation}
	\mathbf{u}(x) = C_1 \, \mathbf{u}_1\, e^{q_1 x} + C_2 \, \mathbf{u}_2 \, e^{q_2 x} + C_3 \, \mathbf{u}_3 \, e^{q_3 x}  
	 							+ C_4 \, \mathbf{u}_4\, e^{q_4 x} + C_5 \, \mathbf{u}_5 \, e^{q_5 x} + C_6 \, \mathbf{u}_6 \, e^{q_6 x} 
	\label{eq_AP01b}	 							
\end{equation}
where $C_j$ are complex coefficients. Among these modes we can distinguish 2 propagating modes of longitudinal waves, 2 propagating modes of bending waves and 2 evanescent/propagating modes of shear waves depending whether the frequency is lower/higher than the cut-off frequency. Associated to each mode, there exist the corresponding wavenumber, which will be denoted by $k_l$ (longitudinal waves), $k_b$ (bending waves) and $k_s$ (shear waves). In Table \ref{tab01} the association between eigenvalues and wave modes is shown. The three first eigenvalues $1 \leq j \leq 3$ will be always linked to the rightwards propagation. 
After some algebra, the characteristic equation $ \det 	\left[\mathbf{A} - q \, \mathbf{I}_6 \right] =0$ can be written as
%
%
\begin{equation}
	\left(q^2 + \kappa_P^2\right) \, \left[    \left(r^2 \, q^2 + \alpha^2 \Omega^2\right)\left(r^2 \, q^2 + \Omega^2\right) -   \alpha^2 \Omega^2\right] = 0
	\label{eq_AP03}
\end{equation}
where
\begin{equation}
	\kappa_P = \omega \sqrt{\frac{\rho A}{EA}} \  , \quad 
	\kappa_S = \omega \sqrt{\frac{\rho A}{GA_z}}  \  , \quad 
	r = \sqrt{\frac{I_y}{A}} \ , \quad
	\alpha = \sqrt{\frac{\rho I_y \, GA_z}{\rho A \, E I_y}} = \frac{\kappa_P}{\kappa_S} 	\  , \quad
	\Omega = \frac{\omega}{\omega_c} \  , \quad
	\omega_c = \sqrt{\frac{GA_z}{\rho I_y}}
\end{equation}
The six roots of the equation are ordered in agreement with the criterium of Table \ref{tab01}, resulting the following frequency-dependent expressions

\begin{align}
	q_1 &= - i  \kappa_P  &
	q_2 &= - i \frac{\Omega}{r} \, \sqrt{\frac{1 + \alpha^2}{2}} \sqrt{\eta + 1} &
	q_3 &= -  \frac{\Omega}{r} \, \sqrt{\frac{1 + \alpha^2}{2}} \sqrt{\eta - 1}     \nonumber \\
	q_4 &= + i   \kappa_P  &
	q_5 &= + i \frac{\Omega}{r} \, \sqrt{\frac{1 + \alpha^2}{2}} \sqrt{\eta + 1}   &
    q_6 &= +  \frac{\Omega}{r} \, \sqrt{\frac{1 + \alpha^2}{2}} \sqrt{ \eta - 1}   
    \label{eq_AP04}
\end{align}
where
\begin{equation}
	\eta = \sqrt{1 + \frac{4 \alpha^2}{(1 + \alpha^2)^2} \frac{1 - \Omega^2}{\Omega^2}}
	\label{eq_AP05}
\end{equation}
The parameter $\eta$ depends in turn on frequency and returns always real values for any frequency, verifying that $\eta \geq 1$ for $\Omega \leq 1$ and $\eta \leq 1$ for $\Omega \geq 1$.  Thus, the complex nature of both eigenvalues $q_3$ and $q_6$ changes depending on the value of $\Omega$. The wavenumbers $k_l, \ k_b$ and $k_s$ of Table \ref{tab01} consistent with the solution obtained in Eq.~\eqref{eq_AP04} are then
\begin{equation}
	k_l = \kappa_P \quad , \quad k_b = \frac{\Omega}{r} \, \sqrt{\frac{1 + \alpha^2}{2}} \sqrt{\eta + 1}  
	\quad , \quad
	k_s = 
	\begin{cases}
			 \displaystyle \frac{\Omega}{r} \, \sqrt{\frac{1 + \alpha^2}{2}} \sqrt{\eta - 1}  & \text{if \ } \omega \leq \omega_c \\
			 \displaystyle \frac{\Omega}{r} \, \sqrt{\frac{1 + \alpha^2}{2}} \sqrt{1 - \eta}  & \text{if \ } \omega \geq \omega_c			 
	\end{cases}
\end{equation}
Modes 1 and 4 represents longitudinal propagating waves. Modes 2 and 5 are bending--flexural modes also with propagating nature. Modes 3 and 6 on the other hand have shear--flexural nature but their propagation depend on the range of frequency. Thus, for low-frequency ($\omega < \omega_c$), they are evanescent and for high-frequency  ($\omega > \omega_c$) they represent shear propagating waves. The eigenvectors are found after plugging the results of Eqs.~\eqref{eq_AP04} into the subspace equation and normalizing. Thus, we find the following set of right- and left-eigenvectors $\mathbf{u}_j, \mathbf{v}_j$ respectively, with the orthogonal relationships
\begin{equation}
	\mathbf{A} \mathbf{u}_j = q_j \, \mathbf{u}_j \ , \quad 
	\mathbf{v}_j^T \, \mathbf{A}  = q_j \, \mathbf{v}_k^T \ , \quad 	
	\mathbf{v}_j^T \, \mathbf{u}_k  = \delta_{jk} \quad 1 \leq j,k \leq 6
	\label{eq_AP08}
\end{equation}
\begin{eqnarray}
			\mathbf{u}_j &=& \displaystyle
				 \begin{cases}
				\displaystyle \frac{1}{\sqrt{2}} \left\{ 1, \ 0, \ 0, \ q_j \, EA, \ 0, \ 0\right\}^T  &  j = 1,4 \\
				\displaystyle \frac{1}{\lambda_j}\left\{0, \ 1, \frac{q_j^2 + \kappa_S^2}{q_j}, \ 0 , \  - \frac{\rho A \omega^2}{q_j}, EI_y (q_j^2 + \kappa_S^2) \right\}^T  & j = 2,3,5,6
			\end{cases} \nonumber \\
			\mathbf{v}_j &=&
			\begin{cases}
				\displaystyle \frac{1}{\sqrt{2}} \left\{ 1, \ 0, \ 0, \frac{1}{q_j \, EA} , \ 0, \ 0\right\}^T &  j = 1,4 \\
				\displaystyle \frac{1}{\lambda_j}\left\{0, \ 1, \frac{q_j}{q_j^2 + \kappa_P^2}, \ 0 , \  - \frac{q_j}{\rho A \omega^2}, \frac{1}{EI_y (q_j^2 + \kappa_P^2)} \right\}^T   
				& j = 2,3,5,6
			\end{cases} 		
			\label{eq_AP10}
\end{eqnarray}
 where
 \begin{equation}
 	\lambda_j  = \sqrt{2 \left(1 + \frac{q_j^2 + \kappa_S^2}{q_j^2 + \kappa_P^2}\right)} \ , \quad j = 2,3,5,6
 	\label{eq_AP09}
 \end{equation}

\section{Green--function of 1D waveguides}
\label{ap_green}

A waveguide governed by the matrix $\mathbf{A}$ with state--vector $\mathbf{u}$ will present in general $6$ modes. Among them, and without loss of generality, the first $3$ modes represent rightwards waves and the last $3$ are leftwards waves. This modes can be either propagating or evanescent modes, depending on the nature of the considered waves as derived in \ref{ap_modes}. For instance, longitudinal waves in rods have modes at each direction. Low frequency flexural waves present 4 modes, 2 are propagating and the other 2 are evanescent. High frequency beam waves (Timoshenko beam) present 4 propagating modes at each direction (bending and shear waves). We will denote by $\mathbf{u}_j$ and $\mathbf{v}_j$ respectively to the right and left 6-components eigenvectors associated to each mode with eigenvalue $q_j$. Thus, we can write
\begin{equation}
	\mathbf{A} \mathbf{u}_j = q_j \, \mathbf{u}_j \ , \quad 
	\mathbf{v}_j^T \, \mathbf{A}  = q_j \, \mathbf{v}_j^T \ , \quad 	
	\mathbf{v}_j^T \, \mathbf{u}_k  = \delta_{jk} \quad 1 \leq j,k \leq 6
	\label{eq066}
\end{equation}
where $\delta_{jl} $ denotes the Kronecker delta. The main aim of this Section is to find an analytical solution of the Green matrix $\mathbf{G}(x)$ of the system, which verifies the equation
\begin{equation}
	\left( \frac{\textrm{d}}{\textrm{d} x} -  \mathbf{A} \right) \mathbf{G} = \mathbf{I} \, \delta (x)
	\label{eq074}
\end{equation}
Let us consider the same problem in vector form, aimed to find $\mathbf{u}(x)$ from the equation
\begin{equation}
	\left( \frac{\textrm{d}}{\textrm{d} x} -  \mathbf{A} \right) \mathbf{u} = \mathbf{Q} \, \delta (x)
	\label{eq075}
\end{equation}
where in principle $\mathbf{Q}$ is a column vector. Considering a point-force excitation located at $\mathbf{q}(x) = \mathbf{Q} \, \delta(x)$, the wavefield at any point of the waveguide can be written in terms of the exponential matrix as
\begin{equation}
	\mathbf{u}(x) = e^{\mathbf{A}(x-x_0)} \, \mathbf{u}(x_0) + \int_{\xi=x_0}^{x}  e^{\mathbf{A}(x-\xi)} \, \mathbf{Q} \,  \delta(\xi) \, d\xi
	= 
	\begin{cases}
		e^{\mathbf{A}(x-x_0)} \, \mathbf{u}(x_0) 		&		\quad  x < 0 \\
		e^{\mathbf{A}(x-x_0)} \, \mathbf{u}(x_0)   +  e^{\mathbf{A}x} \, \mathbf{Q} & 	\quad  	\quad  x \geq 0
	\end{cases}								
	\label{eq067} 
\end{equation}
where $x_0<0$ is any point to the left of the origin. On the other side, the wavefield solution radiates waves in both directions, rightwards for $x>0$ and leftwards for $x<0$. Therefore the solution should have the following structure
\begin{equation}
	\mathbf{u}(x) = 
	\begin{cases}
						 	 -   C_4 \, \mathbf{u}_4\, e^{q_4 x}  - C_5 \mathbf{u}_5 \, e^{q_5 x} -  C_6 \mathbf{u}_6 \, e^{q_6 x}  & \quad x < 0 \\
		\phantom{-} C_1 \, \mathbf{u}_1\, e^{q_1 x} + C_2 \mathbf{u}_2 \, e^{q_2 x} + C_3 \mathbf{u}_3 \, e^{q_3 x}  & \quad x \geq 0 
		\label{eq069}						
	\end{cases}
\end{equation}
where the coefficients $C_j, \ 1 \leq j \leq 6$ are unknown to be obtained. Evaluating at $x=0^-$ and at $x=0^+$ in both Eqs.~\eqref{eq067} and \eqref{eq069} we have
\begin{eqnarray}
	 e^{-\mathbf{A}x_0}  \mathbf{u}(x_0) 	 						   & = & -\sum_{j > 3} C_j \, \mathbf{u}_j  \ , \qquad (x=0^-) 	\label{eq068a}\\
	 e^{-\mathbf{A}x_0}  \mathbf{u}(x_0)  + \mathbf{Q}	 & = & \phantom{-}\sum_{j \leq 3} C_j \, \mathbf{u}_j    \ , \qquad (x=0^+)
	\label{eq068b}
\end{eqnarray}
The above equations represent a system of $12$ linear equations with $12$ unknowns: $6$ components of $\mathbf{u}(x_0)$ and the $6$ coefficients $C_j, \ 1 \leq j \leq 6$. Substracting Eq.~\eqref{eq068b} and Eq.~\eqref{eq068a}
\begin{equation}
\mathbf{Q} = 	\sum_{j =1 }^{6} C_j \, \mathbf{u}_j 
	\label{eq070}
\end{equation}
Premultipling now by $\mathbf{v}_l^T$ and using the orthogonal relationships we can obtain the explicit value of each coefficient $C_l$ as
\begin{equation}
	C_l  = \mathbf{v}_l^T \mathbf{Q} 	\quad , \quad 1 \leq l \leq 6
	\label{eq071}
\end{equation}
Now, plugging the above result into Eq.~\eqref{eq069} we find and expression with the form
\begin{equation}
	\mathbf{u}(x) = \mathbf{G}(x) \, \mathbf{Q}
	\label{eq072}
\end{equation}
where the matrix of Green functions $\mathbf{G}(x)$ is finally 
\begin{equation}
	\mathbf{G}(x) = 
	\begin{cases}
		\displaystyle \phantom{-}\sum_{j =1}^{3}  \mathbf{u}_j  \, \mathbf{v}_j^T  \, e^{q_j x}  & \quad x \geq 0 \\
		\displaystyle 	-\sum_{j = 4}^{6}  \, \mathbf{u}_j  \, \mathbf{v}_j^T   \, e^{q_j x}  & \quad x < 0
	\end{cases}
	\label{eq073}						
\end{equation}
Using the orthogonal relationships it is straightforward that the so obtained Green function verifies the following properties
\begin{itemize}
	\item [(i)] $\mathbf{G}'(x) = \mathbf{A} \, \mathbf{G}(x)$ for $x \neq 0$ 
	\item[(ii)] $\mathbf{G}(0^+) - \mathbf{G}(0^-) = \mathbf{I}_{6}$
\end{itemize}
The second property (ii) is consistent since the matrix $\mathbf{G}(x)$ relates the components of $\mathbf{u}(x)$ with their-selves, therefore the main diagonal captures the discontinuity jumps in the field. Thus an external point-force produces a jump in the field of sectional forces. Likewise for the rest of components of the state-vector.

\section*{Acknowledgments}

M.L. and V.R.-G. are grateful for the partial support under Grant No. PID2020-112759GB-I00 funded by 
MCIN/AEI/10.13039/501100011033. V.R.-G. acknowledge support from 
Grant No. CIAICO/2022/052 of the ``Programa para la promoci\'on de la investigaci\'on cient\'ifica, el desarrollo tecnol\'ogico y la innovaci\'on en la Comunitat Valenciana'' funded by Generalitat Valenciana. M.L is grateful for support under the ``Programa de Recualificaci\'on del Sistema Universitario Espa\~nol para 2021-2023'', (funded by ``Instrumento Europeo de Recuperaci\'on (Next Generation EU) en el marco del Plan de Recuperaci\'on, Transformaci\'on y Resiliencia de Espa\~na'', a trav\'es del Ministerio de Universidades.

\section*{References}
\bibliographystyle{elsarticle-num} 
\bibliography{bibliography}





\end{document}